\documentclass[fleqn,usenatbib]{mnras}

\usepackage{newtxtext,newtxmath}

\usepackage[T1]{fontenc}

\DeclareRobustCommand{\VAN}[3]{#2}
\let\VANthebibliography\thebibliography
\def\thebibliography{\DeclareRobustCommand{\VAN}[3]{##3}\VANthebibliography}

\usepackage{graphicx}	
\usepackage{amsmath}	

\usepackage{amssymb}	
\usepackage{gensymb}	

\newcommand{\Mpc}{\mbox{$\>{\rm Mpc}$}}
\newcommand{\kpc}{\mbox{$\>{\rm kpc}$}}
\newcommand{\pc}{\mbox{$\>{\rm pc}$}}
\newcommand{\Gyr}{\mbox{$\>{\rm Gyr}$}}

\newcommand{\Msun}{\>{\rm M_{\odot}}}

\newcommand\degrees{^\circ}

\newcommand{\Mstar}{\mbox{$M_{\star}$}}
\newcommand{\logMstar}{\mbox{$\log(M_{\star}/\mathrm{M}_{\odot})$}}
\newcommand{\logMT}{\mbox{$\log(M_T/\mathrm{M}_{\odot})$}}
\newcommand{\Reff}{\mbox{$R_{\rm eff}$}}
\newcommand{\Krot}{\mbox{$K_{\rm rot}$}}
\newcommand{\Abar}{\mbox{$A_{\rm bar}$}}
\newcommand{\Atwomax}{\mbox{$a_{2,\mathrm{max}}$}}
\newcommand{\Rbar}{\mbox{$R_{\rm bar}$}}
\newcommand{\RbarReff}{\mbox{$R_{\rm bar}/R_{\rm{eff}}$}}

\newcommand{\fbp}{\mbox{$f_{\rm BP}$}}
\newcommand{\tbp}{\mbox{$t_{\rm BP}$}}
\newcommand{\tbar}{\mbox{$t_{\rm bar}$}}
\newcommand{\tbuck}{\mbox{$t_{\rm buck}$}}
\newcommand{\thf}{\mbox{$t_{\rm 1/2}$}}
\newcommand{\zbar}{\mbox{$z_{\rm bar}$}}
\newcommand{\zbp}{\mbox{$z_{\rm BP}$}}
\newcommand{\zbuck}{\mbox{$z_{\rm buck}$}}
\newcommand{\zhf}{\mbox{$z_{1/2}$}}
\newcommand{\facc}{\mbox{$f_{\rm acc}$}}
\newcommand{\sigzsigR}{\mbox{$\sigma_z/\sigma_R$}}
\newcommand{\sigzsigRinbar}{\mbox{$(\sigma_z/\sigma_R)_{\mathrm{inbar}}$}}

\newcommand{\zrms}{\mbox{$z_{\rm rms}$}}
\newcommand{\zrmsReff}{\mbox{$z_{\rm rms}/R_\mathrm{eff}$}}
\newcommand{\fginbar}{\mbox{$f_{g\rm ,inbar}$}}
\newcommand{\sigzinbarvc}{\mbox{$\sigma_{z,{\mathrm{inbar}}}/V_\mathrm{c}$}}
\newcommand{\sigRinbarvc}{\mbox{$\sigma_{R,{\mathrm{inbar}}}/V_\mathrm{c}$}}

\newcommand{\numdiscs}{\mbox{$608$}}
\newcommand{\numbars}{\mbox{$191$}}
\newcommand{\numnonBPbars}{\mbox{$85$}}
\newcommand{\numbarsinclshort}{\mbox{$266$}}
\newcommand{\numnonbars}{\mbox{$342$}}
\newcommand{\numbps}{\mbox{$106$}}
\newcommand{\numbuck}{\mbox{$52$}}
\newcommand{\numnonbuck}{\mbox{$54$}}
\newcommand{\Bcrit}{\mbox{$0.040$}}
\newcommand{\errortext}{The shaded areas represent the $95$ per cent confidence interval about the median from bootstrap resampling.}

\usepackage[dvipsnames]{xcolor}
\definecolor{dgreen}{RGB}{0,150,40}

\def\eg{{\it e.g.}}

\def\ie{{\it i.e.}}

\def\araa{ARAA}
\def\aj{AJ}
\def\apj{ApJ}
\def\apjl{ApJ}
\def\apjs{ApJS}
\def\aap{A\&A}
\def\mnras{MNRAS}
\def\pasp{PASP} 
\def\nat{Nature}

\def\aaps{A\&AS}

\long\def\Ignore#1{\relax}

\defcitealias{erwin_debattista17}{ED17}

\title[BP bulges in TNG50]{The interplay between accretion, downsizing and the formation of box/peanut bulges in TNG50}

\author[Anderson et al.]{Stuart Robert Anderson,$^{1}$\thanks{E-mail: sra.ngc1300@gmail.com} Steven Gough-Kelly,$^{1}$ Victor P. Debattista,$^{1,2}$ Min Du,$^{3}$
\newauthor{Peter Erwin,$^{4,5}$ Virginia Cuomo,$^6$ Joseph Caruana,$^{7,2}$ Lars Hernquist$^8$}
\newauthor{and Mark Vogelsberger$^9$}
\\
$^{1}$Jeremiah Horrocks Institute, University of Central Lancashire, Preston, PR1 2HE, UK\\
$^{2}$Institute of Space Sciences \& Astronomy, University of Malta, Msida MSD 2080, Malta\\
$^{3}$Department of Astronomy, Xiamen University, Xiamen, Fujian 361005, China\\
$^{4}$Max-Planck-Insitut f\"{u}r extraterrestrische Physik, Giessenbachstrasse, 85748 Garching, Germany \\
$^{5}$Universit\"{a}ts-Sternwarte M\"{u}nchen, Scheinerstrasse 1, D-81679 M\"{u}nchen, Germany \\
$^{6}$Instituto de Astronom\'ia y Ciencias Planetarias, Universidad de Atacama, Avenida Copayapu 485, 1350000 Copiap\'o, Chile\\
$^{7}$Department of Physics, Faculty of Science, University of Malta, Msida MSD 2080, Malta\\
$^{8}$Harvard–Smithsonian Center for Astrophysics, 60 Garden Street, Cambridge, MA 02138, USA\\
$^{9}$Department of Physics and Kavli Institute for Astrophysics and Space Research, Massachusetts Institute of Technology, Cambridge, MA 02139, USA\\
}

\date{Accepted XXX. Received YYY; in original form ZZZ}

\pubyear{2023}

\begin{document}
\label{firstpage}
\pagerange{\pageref{firstpage}--\pageref{lastpage}}
\maketitle

\begin{abstract}
From the TNG50 cosmological simulation we build a sample of 191
well-resolved barred galaxies with stellar mass $\logMstar >
10$ at $z=0$. We search for box/peanut bulges (BPs) in this sample, finding them
in 55 per cent of cases. We compute $\fbp$, the BP probability for barred
galaxies as a function of $\Mstar$, and find that this rises to a
plateau, as found in observations of nearby galaxies. The transition
mass where $\fbp$ reaches half the plateau value is $\logMstar = 10.13\pm0.07$,
consistent with the observational value within measurement
errors. We show that this transition in $\fbp$ can be attributed to
the youth of the bars at low $\Mstar$, which is a consequence of
downsizing. Young bars, being generally shorter and
weaker, have not yet had time to form BPs. At high mass, while we find
a plateau, the value is at $\sim 60$ per cent whereas observations saturate
at $100$ per cent. We attribute this difference to excessive heating in
TNG50, due to merger activity and to numerical resolution
effects. BPs in TNG50 tend to occur in galaxies with more
quiescent merger histories. As a result, the main driver of whether a
bar hosts a BP in TNG50 is not the galaxy mass, but how long and
strong the bar is. Separating the BP sample into those that have
visibly buckled and those that have not, we find that fully half of BP
galaxies show clear signs of buckling, despite the excessive heating
and limited vertical resolution of TNG50.

\end{abstract}

\begin{keywords}
galaxies: bulges -- galaxies: evolution -- galaxies: kinematics and dynamics -- galaxies: structure
\end{keywords}


\section{Introduction}
\label{s:intro}

Box/peanut bulges (BPs) are the vertically extended inner portions of some galactic bars, with a characteristic `X' shape when viewed side-on. Bars are present in most disc galaxies \citep{eskridge_etal_00, menendez-delmestre+07, marinova_jogee_07, Erwin_2018, Erwin_2019}, and are key drivers of much of their secular evolution \citep{weinberg85, debattista_sellwood98, athanassoula02, athanassoula_misiriotis02, athanassoula03, kormendy_kennicutt04, hol_etal_05, debattista+06, ceverino_klypin_07, dubinksi+2009}. The Milky Way (MW) also has a BP \citep[][and references therein]{nataf+10, mcwilliam_zoccali10}. Therefore an understanding of galaxy evolution, including of the MW, requires a detailed understanding of the conditions under which BPs form.

Observationally, \citet[][hereafter ED17]{erwin_debattista17} studied 84 local barred galaxies from the Spitzer Survey of Stellar Structure in Galaxies \citep[S$^4$G,][]{sheth+2010}, finding the fraction of BPs amongst barred galaxies to be a strong function of stellar mass, and that above a stellar mass $\logMstar \simeq 10.4$ the majority of barred galaxies host BPs. This was updated with a larger dataset of 131 galaxies by Erwin, Debattista \& Anderson (submitted; hereafter EDA23), yielding a turning point of $\logMstar \simeq 10.3\pm0.05$. Using kinematic signatures, \citet{gadotti+2020} found a fraction of 62\% BPs in a sample of 21 massive barred galaxies from the MUSE TIMER project. \citet{marchuk_etal_2022} also found an upturn in BP fraction at $\logMstar \simeq 10.4$ in their sample of 483 edge-on galaxies from the Dark Energy Spectroscopic Instrument survey, albeit with a much lower fraction overall.

BPs have been studied in detail in simulations \citep[e.g.][]{raha+91, athanassoula_misiriotis02, berentzen+98, bureau_athanassoula05, martinez-valpuesta+06, saha_gerhard13, fragkoudi+17, debattista+17, saha+2018, lokas_2019, ciambur+2021}. They show that one formation mechanism for BPs is the buckling instability \citep{raha+91, merritt_sellwood94, martinez-valpuesta_shlosman04, smirnov_sotnikova19}, a sudden asymmetric bending of the bar in its inner regions. Observations have recognized a few ($\lesssim 10$) galaxies currently undergoing buckling \citep{erwin_debattista16,li+17,xiang+2021}. Strong buckling produces long-lasting asymmetries about the mid-plane, which may be recognized for a few Gyr in edge-on galaxies, but these have not yet been detected \citep{cuomo_etal_2022}. An alternative BP formation mechanism, the trapping of stellar orbits into vertical resonances, is more gradual \citep{combes_sanders81, combes+90, quillen02, sellwood_gerhard_2020}. What fraction of the observed BPs is produced by the two mechanisms is as yet unknown.

Large-volume galaxy-formation simulations are now reaching resolutions where they can begin to capture realistic secular evolution for large samples of galaxies \citep{vogelsberger+2020}. While many studies of BP bulges have been in isolated simulations, there have been relatively few using cosmological simulations. \citet{buck+18} studied a Milky Way-like galaxy from the NIHAO high-resolution zoom-in simulations, focusing on the BP. \citet{debattista+19} studied a strongly barred galaxy with an `X' shape from the FIRE cosmological suite, and found that the density bimodality of the `X'-shape decreases in strength with stellar age. \citet{gargiulo+2019} examined the 30 zoom-in simulations in the Auriga suite \citep{grand+17,grand+19} specifically designed to replicate Milky Way-like galaxies, finding at least three with BPs at $z=0$. \citet{blazquez-calero_etal_2020} examined 21 barred galaxies in the Auriga simulation and found four galaxies with BPs ($19$ per cent) and an additional two which were buckling at $z=0$. Using the same simulations, \citet{fragkoudi+2020} found five galaxies with BPs, and four with weak BPs, for a total fraction of $\sim30$ per cent of barred galaxies, considerably lower than in observations.

In this study, we use the TNG50 simulation \citep{nelson_etal_2019,nelson+19,pillepich+19} to study galaxies with BPs at $z=0$. TNG50 is the highest resolution run of the IllustrisTNG project \citep{vogelsberger+2014, vogelsberger_nature_2014,springel+18,nelson+2018,naiman+18,marinacci+18} to date, and offers an unrivalled opportunity to investigate BP galaxies with reasonably high resolution. This simulation enables us to study a significantly larger self-consistent sample of BP galaxies than has previously been possible.
The aim of this study is to provide insight into the conditions which lead to the formation of BPs with a statistically significant sample. %
We examine the BP fraction amongst barred galaxies, and discern BPs formed by strong buckling, and those formed by weak buckling or resonant trapping. We compare our findings to observations of the local universe, analysing and offering explanations for any differences.

This paper is structured as follows. In Section~\ref{s:TNG50_overview}, we briefly introduce the TNG50 simulation. The methods used to detect and measure bars and BPs are variations of techniques used previously, and are presented in Appendix~\ref{s:sample_selection}. In Section~\ref{s:stats_z0} we present the fraction of BPs as a function of stellar mass in TNG50. In Section~\ref{s:logistic_regression_fits} we model the BP fraction as a function of various galactic characteristics, and in Section~\ref{s:kinematics} we analyse the differences in evolution between the galaxies with BPs at $z=0$ and those without. In Section~\ref{s:downsizing} we examine the impact of bar age on the frequency of BPs, and in Section~\ref{s:gas} we analyse the role of gas. We discuss the implications of our findings in Section~\ref{s:discussion}, and summarise in Section~\ref{s:summary}. The reader interested in measurement details may wish to read the Appendix right after Section~\ref{s:TNG50_overview}.


\section{TNG50}
\label{s:TNG50_overview}

IllustrisTNG \footnote{\url{https://www.tng-project.org}} \citep{vogelsberger+2014,vogelsberger_nature_2014,springel+18,nelson+2018,naiman+18,marinacci+18} is a suite of advanced cosmological simulations within the $\Lambda$CDM framework that employs magneto-hydrodynamics as well as a dual-mode thermal and kinematic black hole feedback model, described in detail in \citet{weinberger_at_al_2017} and \citet{pillepich+18_2}. The free parameters of the model were chosen to reproduce observed integrated trends of galactic properties.

We use the highest resolution run of the IllustrisTNG suite, TNG50, which simulated $2\times 2160^3$ dark-matter particles and gas cells within a volume of side-length $35/h$ or $\sim 50$ c$\Mpc$. Alongside a baryonic mass resolution of $8.5 \times 10^4 \Msun$ and gravitational softening length for stellar particles of $\epsilon = 288 \pc$ ($1\geq z \ge0$), $\epsilon\sim500 \pc$ ($z>1$), the simulation also achieves gas cell sizes as small as $70 \pc $ in dense star-forming regions. This creates large statistical samples of galaxies at `zoom'-like resolution \citep{nelson+19,pillepich+19}. 

The IllustrisTNG Collaboration identified dark matter halos within the simulation at each timestep using the friends-of-friends algorithm \citep{davis+1985}; subhalos within each halo were identified using the \texttt{SUBFIND} algorithm \citep{springel+01, dolag+09}. Galaxies are defined as gravitationally bound stellar masses within a halo or subhalo. In conducting our analysis, we extract all the bound particles (gas, stellar, and dark matter particles) within a (sub)halo associated with a galaxy and place the origin at the centre of the potential. We then rotate each galaxy by aligning the angular momentum vector within $2R_{\rm eff}$ of the stellar disc with the $z$-axis, resulting in the stellar disc being in the ($x, y$)-plane. We extract the main leaf progenitor branch \citep{rodriguez-gomez+15} for each of our sample galaxies.

While TNG50 has a large number of galaxies modelled with state-of-the-art subgrid physics, none the less it has some systematic effects which limit the analysis we can employ. In particular, at $z=0$, TNG50 has a stellar force softening length of $288 \pc$. This means that at small distances, the force resolution is comparable to the disc thickness, and bending waves (such as those involved in the buckling instability) are not well captured  \citep{merritt_sellwood94}. Hence dynamical processes are not fully resolved for small bars, and we remove galaxies with small bars ($\Rbar < 2.6 \kpc$) from our analysis (see Section~\ref{ss:bar_sample} for a detailed discussion).

We emphasize that we only seek galaxies with BPs at the \emph{current epoch}, examining their evolution since $z=2$. We do not consider galaxies which may have had BPs at earlier epochs but do not at $z=0$. To ensure enough stellar particles for our study, we analyse galaxies with $\logMstar\ge10.0$. The methods used to find disc galaxies, those with bars at $z=0$ and those barred galaxies with BPs at $z=0$, are all described in detail in Appendix~\ref{s:sample_selection}. Briefly, our initial sample consists of $\numdiscs$ disc galaxies. We analyse only bars with radius $\ge2.6 \kpc$, justifying this in Section~\ref{ss:refine_rbar_cut}. To compare with galaxies without BPs, we construct a sample of `control' galaxies from the barred non-BP sample. We match a non-BP galaxy to a BP galaxy by stellar mass, with replacement ($\ie$ the control galaxy can be matched again with another BP), producing the `Control' sample. We denote those BP galaxies which experienced strong buckling as the `BCK' sample, and those which experienced weak or no buckling as the `WNB' sample. The method used to differentiate between these classifications is described in Section~\ref{ss:assessment_buckling}.

For the reader's convenience, a list of the notation we use for parameters of the galaxies is given in Table~\ref{tab:symbols}. Most are calculated within a sphere of radius $10R_\mathrm{eff}$. However, the kinematics and heights are computed within a cylindrical annulus spanning $\Reff<R<2\Reff$ so that we avoid the centre and any warping in the outer disc, although this does not always completely exclude the entire bar radius. This gives a fair comparison between those galaxies destined to host BPs, and those destined not to. The reader interested only in our results can consult Table~\ref{tab:symbols} and skip Appendix~\ref{s:sample_selection}.

\begin{table}
\centering
\caption{Summary of notation used in the paper.}
\label{tab:symbols}
\begin{tabular}{ll}

\hline
Symbol & Description\\
\hline
$\Reff$  & Cylindrical radius containing half the total stellar mass\\

$\Mstar$ & Stellar mass of a galaxy (within $r \leq 10R_\mathrm{eff}$)\\

$M_g$ & Gas mass of a galaxy (within $r \leq 10R_\mathrm{eff}$) \\
$\Atwomax$ & Bar strength, calculated as described in Section~\ref{ss:bar_sample}\\

$A_{\mathrm{bar}}$ & Global bar strength (within $r \leq 10R_\mathrm{eff}$), \\ & calculated via the $m=2$ Fourier amplitude, \\
    & as described in Section~\ref{ss:bar_sample} \\

$\Rbar$ & Bar radial extent, calculated as described in Section~\ref{ss:bar_sample}\\

$f_g$ & The fraction of baryonic mass in gas, $M_g/(\Mstar + M_g)$ \\

$f_{g, \mathrm{inbar}}$ & The fraction of baryonic mass in gas within the cylindrical\\
            & radius $\Rbar$\\

$\Krot$ & Measure of the fraction of kinetic energy attributed to \\
        & rotational motion, defined in Section~\ref{ss:bar_sample}\\

$\zrms$ & The root mean square mass-weighted height \\
           & (computed within $\Reff<R<2\Reff$) \\

$\sigma_R$, $\sigma_\phi$, $\sigma_z$  & The dispersion in the radial, azimuthal and vertical velocities\\
 & (within $r<10 \Reff$) \\

$\mathcal{B}$ & `Strength' of the BP bulge, measured as described in \\
            & Section~\ref{ss:BP_strength}\\

$\facc$ & Fractional accreted stellar mass of a galaxy. A           stellar
    \\ & particle is considered to have been formed in situ if the \\ & galaxy in which it formed lies along the main progenitor\\ & branch (in the merger trees) of the galaxy in which it \\ &  is currently found. Otherwise, it is tagged as ex situ, \\ & \ie\ accreted \citep{rodriguez-gomez+15} \\

$\mathcal{F}$ & $M_{\rm{gal}}(z)/M_{\rm{group}}(z)$, where      $M_{\rm{gal}}(z)$ is the total mass of \\
    & the galaxy (dark matter + gas and stars), and $M_{\rm{group}}(z)$ is \\
    & the total mass of the friends of friends group in which the \\
    & galaxy is located at each redshift $z$ \\

$\tbar (\zbar)$ & Time (redshift) of bar formation, assessed as described\\
    & in Section~\ref{ss:assessment_t_bar} \\

$\tbp (\zbp)$ & Time (redshift) of BP formation, assessed as described\\
    & in Section~\ref{ss:time_BP_formation} \\

$\tbuck (\zbuck)$ & Time (redshift) of buckling, assessed as described\\
    & in Section~\ref{ss:assessment_buckling} \\

$\thf (\zhf)$ & Time (redshift) at which the galaxy's stellar mass \\
                    & (within a sphere of radius $10R_\mathrm{eff}$) reaches half its value \\
                    & at $z=0$\\

$V_\mathrm{c}$  & The circular velocity at a cylindrical radius of $2\Reff$ \\

$M_T$ & The transition mass in the BP fraction, where the \\
    & fraction reaches $50$ per cent of its asymptotic value at high\\
    & mass\\

\hline
\end{tabular}\\
\end{table}

\section{BP frequency at redshift zero}
\label{s:stats_z0}

From our sample of $\numdiscs$ disc galaxies with $\logMstar\ge10.0$, we find $\numbars$ (32\%) have bars with $\Rbar\ge 2.6\kpc$ at $z=0$ (Section~\ref{ss:bar_sample}), which is the `barred sample'. Of these, $\numbps$ (55\%) have BPs at $z=0$ (the `BP sample'). In the BP sample, $\numbuck$ ($49$ per cent) have strongly buckled in the past and $\numnonbuck$ ($51$ per cent) have never strongly buckled. We find one galaxy (ID 608386) which appears to be in the process of buckling at $z=0$. For this work we consider this galaxy to be a non-BP galaxy but we reject it as a control (Section~\ref{ss:control_sample} describes how BP galaxies are assigned controls). Although not the focus of this study, we find five galaxies that appear to have buckled but do not have a BP at $z=0$, and another two galaxies which did not buckle strongly, but had a BP that is no longer discernible by $z=0$. A detailed analysis of these galaxies is outside the scope of this study. We summarise these statistics in Table~\ref{tab:z0_stats}.

\begin{table}
\label{tab:stats}
\centering
\caption{\label{tab:z0_stats}Sample sizes of TNG50 galaxy classifications.}
\begin{tabular}{ll}

\hline
Classification & Number\\
\hline
Disc galaxies & $\numdiscs$\\
Barred galaxies ($\Rbar\ge 2.6\kpc$) & $\numbars$\\
\hspace{3mm} BP galaxies & $\numbps$\\
\hspace{6mm} Strongly buckling galaxies (BCK) & $\numbuck$\\
\hspace{6mm} Weak/non-buckling galaxies (WNB) & $\numnonbuck$\\

\hline
\end{tabular}\\
\end{table}

\begin{figure}
  \includegraphics[width=\hsize]{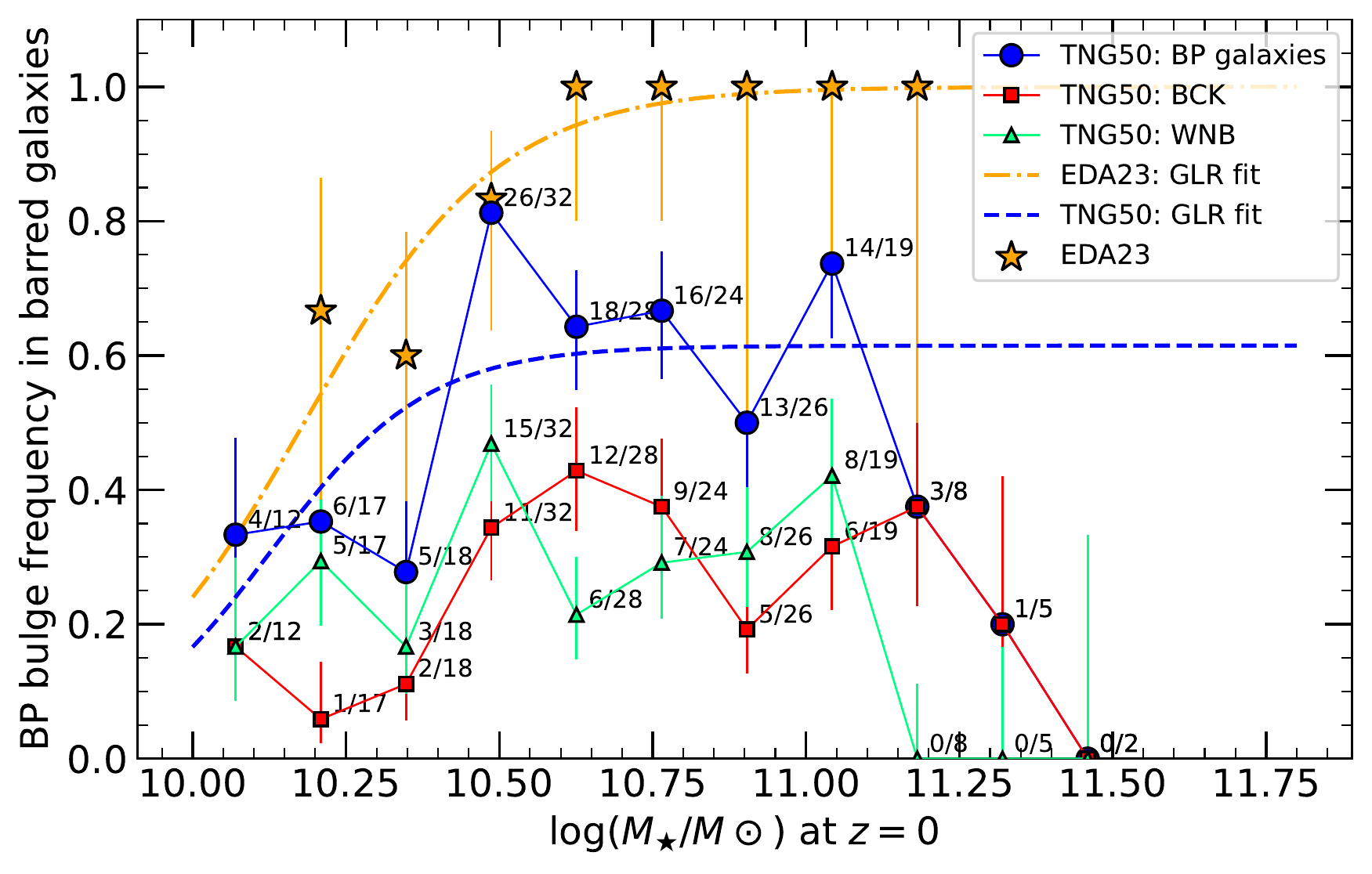}
  \caption{Dependence of the BP fraction on stellar mass at $z=0$ for the sample of $\numbars$ barred galaxies (solid blue). It is split into the BCK (red) and WNB (green) populations. The fractions next to each point show the number of BP galaxies/number of barred galaxies in each mass bin.
  The blue dashed line represents a generalized logistic regression (GLR) fit to the data for all TNG50 BPs (see text for details). The orange stars and dot dashed line represent the logistic regression fit to the observational data of EDA23 (restricted to bars of radius $\ge 2.6$ kpc to match our selection, and limited to the same mass range as in this paper's barred sample).
  Error bars are the $68$ per cent (1$\sigma$) confidence limits from the \citet{wilson_1927} binomial confidence interval. TNG50 appears to significantly under-produce BP bulges, particularly at higher mass.}
  \label{fig:z0BPfracdist}
\end{figure}

Fig.~\ref{fig:z0BPfracdist} shows the variation of the BP fraction amongst barred galaxies ($\fbp$) as a function of stellar mass at $z=0$ in TNG50. \citetalias{erwin_debattista17} found an increasing $\fbp$ with $\Mstar$ (see their figure~5) from an analysis of 84 local barred galaxies in the S$^4$G catalog \citep{sheth+2010}. We also show results from EDA23 which uses a larger dataset (131 galaxies) from a distance-limited subset of S$^4$G, using {\it Spitzer} 3.6-micron images. From the EDA23 sample, we exclude bars with radii $<2.6$ kpc and restrict the mass range to our TNG50 barred sample. This cut results in 27 galaxies, 23 of which ($85$ per cent) have BPs.

At low mass ($\logMstar\sim10.1$), the BP fraction in TNG50 is low, similar to the results of \citetalias{erwin_debattista17}.
In TNG50, $\fbp$ increases rapidly for increasing $\Mstar$ until the fraction reaches $\sim 0.6$ at $\logMstar\sim10.5$, followed by a plateau at higher masses. TNG50 does not match the observed BP fraction for $\logMstar \gtrsim 10.6$. All the galaxies with $\logMstar\gtrsim11.2$ in the sample of \citetalias{erwin_debattista17}, and all the galaxies with $\logMstar\gtrsim10.6$ in EDA23, have BPs.

Following \citetalias{erwin_debattista17}, we quantify the mass dependence of the BP fraction using logistic regression, which models whether a barred galaxy has a BP or not, with a function for the probability of a binomial property. We generalize the fitted function to account for the fact that at higher mass, $\fbp$ does not asymptote to 1 in TNG50:
\begin{equation} \label{eq:gen_logistic_regression}
\displaystyle
P(\mu)=\frac{K}{1 + e^{-(\alpha+\beta \mu)}},
\end{equation}
where $\mu\equiv\logMstar$ (or other variable being fitted) and $K$ is the asymptotic fraction of BPs at high mass. $\beta$ is the steepness of the curve, the rate at which the probability increases or decreases with $\mu$. The ratio $-\alpha/\beta$ is the midpoint of the curve, \ie\ the value of $\mu$ where $P(\mu)$ reaches $K/2$.

The procedure fits all data points (\ie\ `has BP? Y/N' indicator versus $\logMstar$ points for all 191 barred galaxies), not the binned data. The parameters of the fit\footnote{The fits were performed using the \texttt{PYTHON} \texttt{SCIPY.OPTIMIZE} library's \texttt{MINIMIZE} function, with a Nelder-Mead solver.} are shown in Table~\ref{tab:logistic_regression}, and the fit is presented in Fig.~\ref{fig:z0BPfracdist} (dashed blue line). In TNG50 the `transition mass' $M_T$, where  $f_\mathrm{BP}$ reaches $50$ per cent of the asymptotic fraction $K$, is $\logMT = 10.13\pm0.07$, which occurs at $\fbp\sim0.31$. \citetalias{erwin_debattista17} found a transition mass of $\logMT \approx 10.37$ at $\fbp=0.5$ ($10.3$--$10.4$ when they considered data from the edge-on galaxies of \citealt{yoshino_yamauchi15}). Note that stellar masses in \citetalias{erwin_debattista17} were determined using estimated stellar M/L ratios, and so have some inherent uncertainty. The logistic regression for the EDA23 dataset, which asymptotes to 1 at high mass, yields a transition mass of $\logMT = 10.19\pm0.06$, a difference of just 0.06 dex from the TNG50 result\footnote{Excluding the $\Rbar$ cut from the EDA23 data, but retaining that in $\logMstar$ (48 galaxies) yields $\logMT = 10.31\pm0.06$.}.

\begin{table*}
\label{tab:logistic_regression}
\centering
\caption{Generalized logistic regression fit to $f_\mathrm{BP}$ as a function of $\mu=\logMstar$ at $z=0$, with $1\sigma$ errors from bootstrap resampling. The functional form fitted is a generalized logistic regression function of the form $f_\mathrm{BP}(\mu) = K/[1+\mathrm{exp}{-(\alpha+\beta \mu)}]$. $M_T$ is the transition mass where $\fbp$ reaches half of the asymptotic fraction at high mass ($K$). We show the generalized fits for all BP galaxies, BCK and WNB samples. The last row shows the logistic regression fits for the EDA23 data, with $\logMstar$ and $\Rbar$ cuts to match those in this study.}

{{\begin{tabular}{lllll}
\hline
Sample & \hspace{5mm} $K$ & \hspace{7mm} $\alpha$ & \hspace{5mm} $\beta$ & $\log(M_T/M_\odot)$\\
\hline
All BPs & $0.61\pm0.05$ & $-79.48\pm17.93$ & $7.85\pm1.70$ & \hspace{2mm} $10.13\pm0.07$\\
BCK & $0.33\pm0.05$ & $-68.50\pm 24.96$ & $6.69\pm 2.35$ & \hspace{2mm} $10.25\pm0.18$\\
WNB & $0.30\pm0.04$ & $-99.08\pm 31.66$ & $9.86\pm 2.40$ & \hspace{2mm} $10.06\pm0.06$\\
EDA23 & $1.00\pm0.00$ & $-64.43\pm24.73$ & $6.33\pm2.30$ & \hspace{2mm} $10.19\pm0.06$\\

\hline
\end{tabular}}}\\
\end{table*}


\section{Multi-variable fits}
\label{s:logistic_regression_fits}

To discern the drivers behind the BP fraction in TNG50 we perform multiple generalized logistic regression fits to equation~\ref{eq:gen_logistic_regression} with the following variables: \Rbar, \tbar, \thf, \sigzsigRinbar, \Atwomax, \logMstar, \facc, \Reff, $f_g$, $f_{g, \mathrm{inbar}}$, \zrms\ and \zrmsReff\ (all defined in Table \ref{tab:symbols}). We also perform the fit with pairs of these variables $(\mu_1,\mu_2)$. In this case we fit:
\begin{equation} \label{eq:gen_logistic_regression_multi}
    \displaystyle
    P(\mu_1, \mu_2)=\frac{K}{1 + e^{-(\alpha+\beta_1 \mu_1+\beta_2 \mu_2)}},
\end{equation}
where $\beta_1$ and $\beta_2$ are the steepness measures of the curve attributed to variables $\mu_1$ and $\mu_2$, respectively.

We assess the power of each variable (and combination of variables) to explain the variation of BP fraction using the Akaike Information Criteria (AIC) \citep{akaike_1974}, which  gives us a basis for comparison\footnote{The AIC for a model is defined as $2k - 2\mathrm{ln}\mathcal{L_\mathrm{max}}$, where $\mathcal{L_\mathrm{max}}$ is the maximized likelihood of the model, and $k$ is the number of parameters in the model.}. AIC compares different models' abilities to explain the greatest amount of variation in the data. We use AIC$_\mathrm{c}$, the AIC corrected for small sample sizes, which is given by $\mathrm{AIC_c}= \mathrm{AIC} + 2k(k+1)/(N-k-1)$ \citep{sugiura78}. In our case, the number of data points is $N=191$, so the correction is small. We change variable $\mu$ in Equation~\ref{eq:gen_logistic_regression} (and combinations of variables $\mu_1$, $\mu_2$ in Equation~\ref{eq:gen_logistic_regression_multi}), and for each variable/combination of variables, compute the AIC$_\mathrm{c}$. The lower the value of the AIC$_\mathrm{c}$, the greater the power of that variable/combination of variables to `explain' whether a galaxy hosts a BP. Differences between fits of $|\Delta\mathrm{AIC_\mathrm{c}}|>6$ are considered strong evidence in favour of the model with lower AIC$_\mathrm{c}$ \citepalias{erwin_debattista17}. Results are shown in Tables \ref{tab:logistic_regression_single_variable} and \ref{tab:logistic_regression_multi_variable}.

\begin{table}
\centering
\caption{Generalized logistic regression fits to $f_\mathrm{BP}$ and AIC$_\mathrm{c}$ as a function of a variety of variables at $z=0$, for the barred sample.}

\label{tab:logistic_regression_single_variable}
{{\begin{tabular}{lllll}
\hline
Variable ($z=0$) & $K$ & \hspace{3mm} $\alpha$ & \hspace{2mm} $\beta$ & \hspace{3mm} AIC$_\mathrm{c}$\\
\hline

\Rbar & $1.0$ & $-6.18$ & $1.78$ & \hspace{2mm} $192.86$ \\
\sigzsigRinbar & $1.0$ & $18.15$ & $-23.37$ & \hspace{2mm} $215.38$ \\
\Atwomax & $1.0$ & $-3.44$ & $9.45$ & \hspace{2mm} $234.52$ \\
\thf & $0.69$ & $10.16$ & $-1.29$ & \hspace{2mm} $247.9$ \\
\facc & $0.63$ & $5.11$ & $-18.05$ & \hspace{2mm} $252.59$ \\
\tbar & $0.64$ & $6.61$ & $-0.58$ & \hspace{2mm} $255.83$ \\
\Mstar & $0.61$ & $-79.48$ & $7.85$ & \hspace{2mm} $261.08$ \\
\zrmsReff & $1.0$ & $1.52$ & $-1.88$ & \hspace{2mm} $261.35$ \\
$f_g$ & $0.59$ & $11.28$ & $-20.0$ & \hspace{2mm} $261.38$ \\
\zrms & $1.0$ & $0.77$ & $-0.17$ & \hspace{2mm} $264.22$ \\
\fginbar & $0.56$ & $635.63$ & $-5000.0$ & \hspace{2mm} $264.88$ \\
\Reff & $0.58$ & $-0.67$ & $1.17$ & \hspace{2mm} $267.41$ \\

\hline
\end{tabular}}}\\
\end{table}

\begin{figure}
  \includegraphics[width=\hsize]{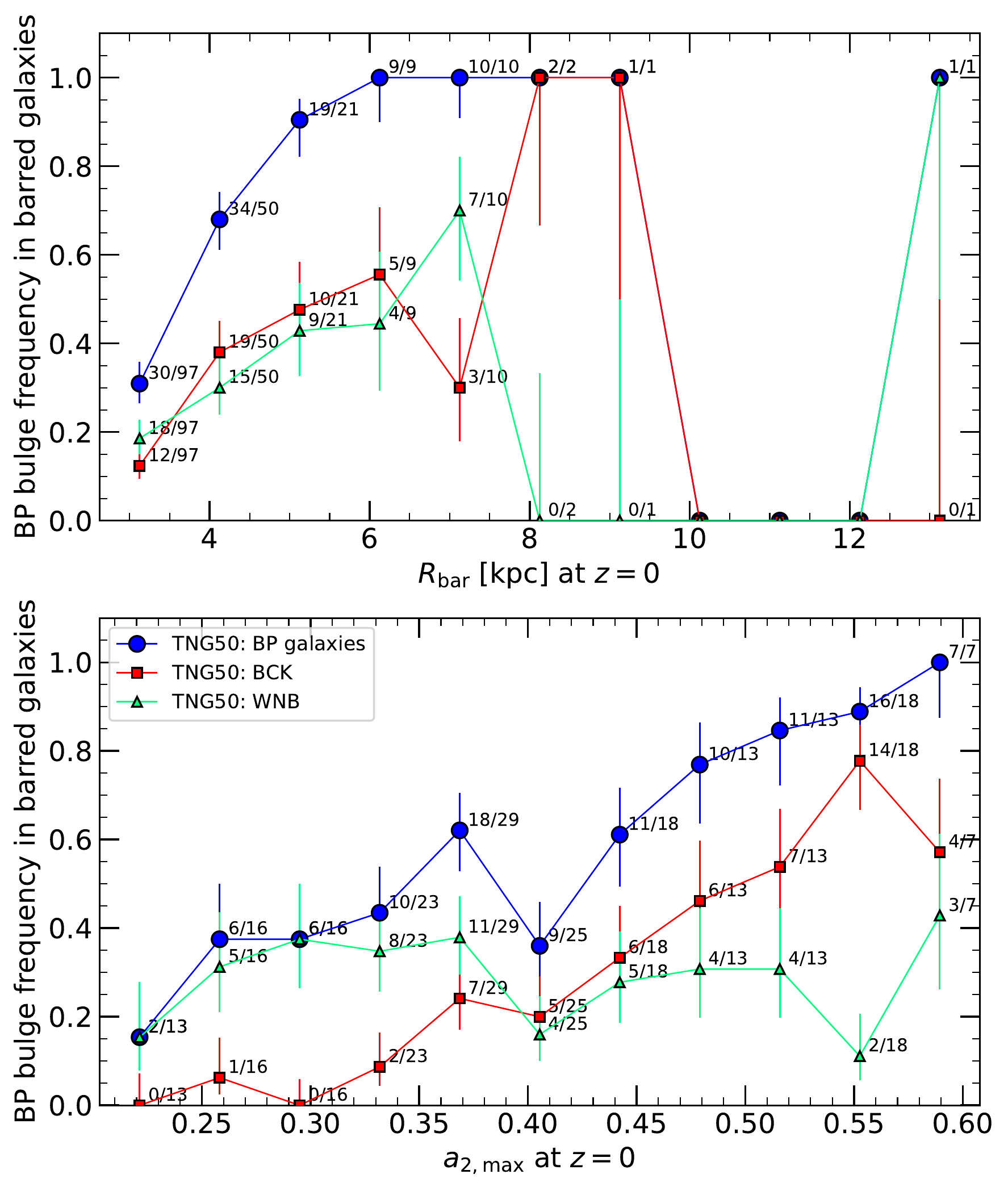}
  \caption{Dependence of the BP fraction on $\Rbar$ (top panel) and $\Atwomax$ (lower panel) at $z=0$ for the sample of $\numbars$ barred galaxies (solid blue). It is split into the BCK (red) and WNB (green) populations. The fractions next to each point show the number of BP galaxies/number of barred galaxies in each mass bin. Error bars are the $68$ per cent (1$\sigma$) confidence limits from the \citet{wilson_1927} binomial confidence interval. The BP fraction in TNG50 is highly dependent on these bar parameters.}
  \label{fig:z0BPfracdistRbarA2max}
\end{figure}

Table \ref{tab:logistic_regression_single_variable} shows that as a single variable, $\Rbar$ has the greatest explanatory power for $\fbp$, rather than $\Mstar$. Table \ref{tab:logistic_regression_multi_variable} shows the results of combinations of $\Rbar$ with other variables. The combinations of $\Rbar$ with $\sigzsigRinbar$ and $\Atwomax$ have the lowest AIC$_\mathrm{c}$, fully $\sim 26$ lower than for $\Rbar$ alone. $\sigzsigRinbar$ is key to the buckling instability, reaching low values before buckling occurs; in this sense it is an inverse proxy for bar strength. We discuss this further in Section~\ref{ss:kinematics_evol}. We also assessed combinations of $\Rbar$ and $\sigzsigRinbar$ with a third variable, and found that the combinations of $\Rbar$, $\sigzsigRinbar$ and $\Atwomax$ gave a slightly lower $\mathrm{AIC}_\mathrm{c}$ but this was not significant ($\Delta\mathrm{AIC_\mathrm{c}} \sim -1$). We conclude that the probability of a bar hosting a BP is a function of bar length and bar strength in TNG50. We show $\fbp$ as a function of $\Rbar$ and $\Atwomax$ in Fig.~\ref{fig:z0BPfracdistRbarA2max}. This clearly shows the strong dependence of BP fraction on both bar parameters -- the stronger, and the longer the bars, the greater the fraction of them which host BPs.

Whereas observations find that $\fbp$ can be fully accounted for by $\logMstar$ \citepalias[also EDA23 who looked at bar strength and found no evidence of an effect]{erwin_debattista17}, in TNG50 $\logMstar$ has relatively poor explanatory power among the variables we have examined. We discuss this disagreement in Section~\ref{s:discussion}.

\begin{table}
\centering
\caption{Generalized logistic regression fits to $f_\mathrm{BP}$ and AIC$_\mathrm{c}$ as a function of combinations of two variables at $z=0$, one of which is $\Rbar$, for the barred sample. $\beta_1$ applies to $\Rbar$, $\beta_2$ to the second variable.}

\label{tab:logistic_regression_multi_variable}
{{\begin{tabular}{llllll}
\hline
Variables ($z=0$) & $K$ & $\alpha$ & \hspace{1mm} $\beta_1$  & \hspace{1mm} $\beta_2$ & \hspace{2mm} AIC$_\mathrm{c}$\\
\hline

\Rbar, \sigzsigRinbar & $1.00$ & $10.43$ & $1.51$ & $-20.44$ & \hspace{2mm} $166.05$ \\
\Rbar, \Atwomax & $1.00$ & $-9.92$ & $1.62$ & $10.89$ & \hspace{2mm} $168.08$ \\
\Rbar, \zrmsReff & $1.00$ & $-4.74$ & $1.85$ & $-2.45$ & \hspace{2mm} $186.89$ \\
\Rbar, \tbar & $1.00$ & $-5.14$ & $1.75$ & $-0.14$ & \hspace{2mm} $190.44$ \\
\Rbar, \fginbar & $1.00$ & $-5.93$ & $1.77$ & $-11.88$ & \hspace{2mm} $191.01$ \\
\Rbar, \facc & $1.00$ & $-5.57$ & $1.71$ & $-3.28$ & \hspace{2mm} $191.33$ \\
\Rbar, \thf & $1.00$ & $-5.01$ & $1.69$ & $-0.13$ & \hspace{2mm} $193.83$ \\
\Rbar, $f_g$ & $1.00$ & $-6.60$ & $1.84$ & $0.81$ & \hspace{2mm} $194.44$ \\
\Rbar, \Mstar & $1.00$ & $-3.04$ & $1.79$ & $-0.30$ & \hspace{2mm} $194.64$ \\
\Rbar, \Reff & $1.00$ & $-6.34$ & $1.79$ & $0.02$ & \hspace{2mm} $194.82$ \\
\Rbar, \zrms & $0.73$ & $-12.67$ & $3.33$ & $1.00$ & \hspace{2mm} $242.97$ \\

\hline
\end{tabular}}}\\
\end{table}

\section{The kinematic state of BP progenitors}
\label{s:kinematics}

The dependence of \fbp\ on galaxy properties in TNG50 barred galaxies motivates us to understand the evolutionary differences between BP and non-BP barred galaxies.
We begin by examining their evolution, showing their median properties at each redshift. Whilst there is variation in the properties for any sample of galaxies, by comparing the medians between different subsamples with the same mass distribution, we can identify key differences in their evolutionary histories. In the following plots, we show the medians and the $95$ per cent confidence intervals about the median (from bootstrap resampling) to highlight trends. We stress that the underlying distributions have significant scatter.

\subsection{Bar size and strength}
\label{ss:bar_characteristics}
\begin{figure*}
  \includegraphics[width=\hsize]{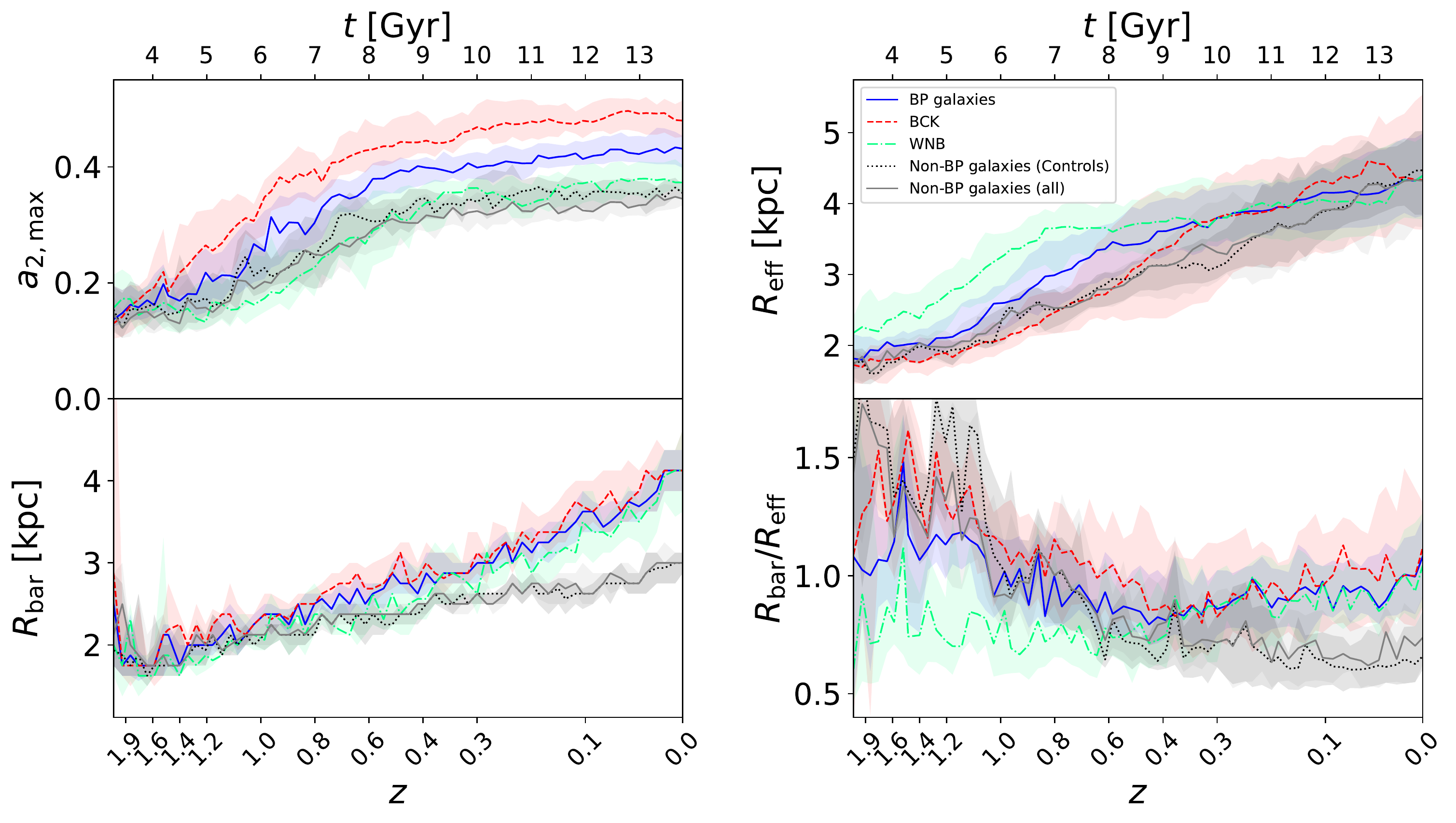}
  \caption{Evolution of the median bar strength (as measured by $\Atwomax$), $\Rbar$, $\Reff$ and $\Rbar/\Reff$ for BP, BCK, WNB, non-BP bars and controls. \errortext}
  \label{fig:evolution_barstrength}
\end{figure*}

We plot the evolution of the bar strength and radius in Fig.~\ref{fig:evolution_barstrength}. This shows that the BP sample has had, since $z\sim1$, stronger bars on average than the non-BP bars. By $z=0$, BP bars are $\sim22$ per cent stronger than those in the controls. At $z=0$, the median bar radius for the BP sample is $4.1 \kpc$, while for the controls it is $3.0 \kpc$. As a fraction of the effective radius, the median $\RbarReff$ is $1.1$ for BP galaxies, but only $0.66$ for the controls, with two-sample Kolmogorov-Smirnov (KS) tests confirming a significant difference in distributions at $z=0$. Thus BPs are found in relatively (and in absolute terms) stronger and longer bars at $z=0$ compared to the bars in the non-BP galaxies. We have checked that there is no monotonic relation between $\Rbar$ and stellar mass for BCK, WNB or non-BP galaxies at $z=0$, at least for our sample with $\Rbar\ge2.6$ kpc (Spearman $\rho\lesssim0.2$). This is consistent with \citet{zhao+2020}, who found that the bar length/stellar mass relation flattens around $\logMstar\sim10.7$ in TNG100. For $\RbarReff$ however, there is a decrease with stellar mass for each subgroup, since $\Reff$ increases with mass. None the less as Fig.~\ref{fig:evolution_barstrength} shows, the finding of BPs being hosted in relatively longer bars holds even when controlling for mass (controls and non-BP galaxies show a similar median value of $\RbarReff$ below that of the BPs).

\begin{figure}
  \includegraphics[width=\hsize]{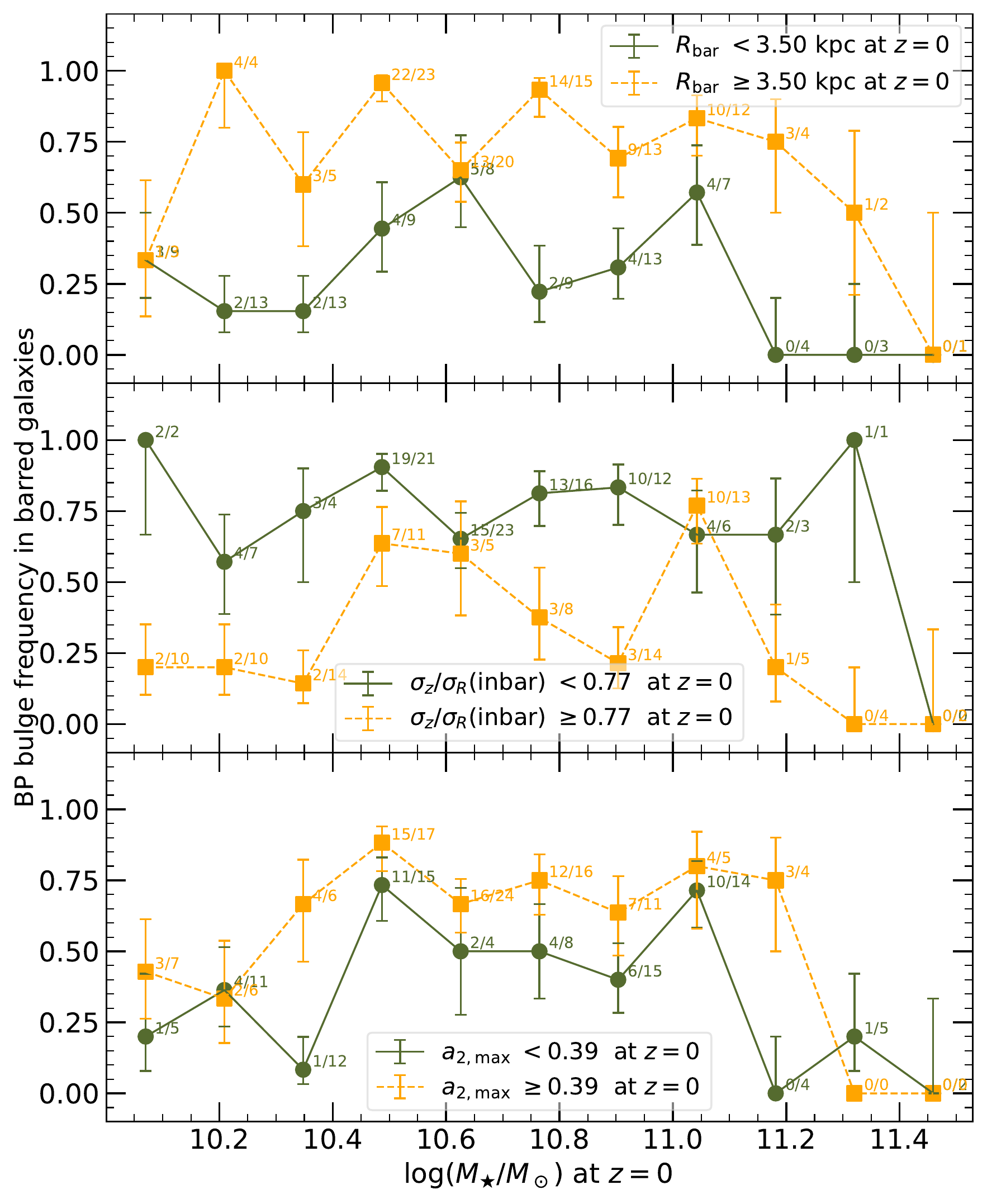}
  \caption{The fraction of barred galaxies hosting BPs at $z=0$, split into high (orange dashed line, squares) and low (olive solid line, circles) populations by median $\Rbar$ (upper panel), median $\sigzsigRinbar$ (middle panel) and median $\Atwomax$ (lower panel). At almost all masses, BPs are hosted preferentially in galaxies whose bars are longer and stronger than the median, and for which $\sigzsigRinbar$ is lower than the median. Error bars are the $68$ per cent (1$\sigma$) confidence limits from the \citet{wilson_1927} binomial confidence interval.}
  \label{fig:BP_frac_by_bar_length_and_strength}
\end{figure}

Fig.~\ref{fig:BP_frac_by_bar_length_and_strength} shows $\fbp$ split at the $z=0$ median bar length ($\Rbar = 3.5\kpc$ top panel), median $\sigzsigRinbar$ (middle panel) and median bar strength ($\Atwomax=0.39$ bottom panel). The denominator in $\fbp$ in this plot in each mass bin is the total number of barred galaxies within the population of short/weak (olive line) and long/strong (orange dashed line) bars. The figure clearly shows a preference, at almost all masses, for BPs to be found in galaxies with longer and stronger bars and lower $\sigzsigRinbar$, consistent with these variables having the lowest AIC$_\mathrm{c}$. As we shall see, however, just because a variable's evolution is different for BPs and non-BPs does not necessarily mean that it will have a low AIC$_\mathrm{c}$.

\subsection{Kinematics and thickness}
\label{ss:kinematics_evol}

\begin{figure}
  \includegraphics[width=\hsize]{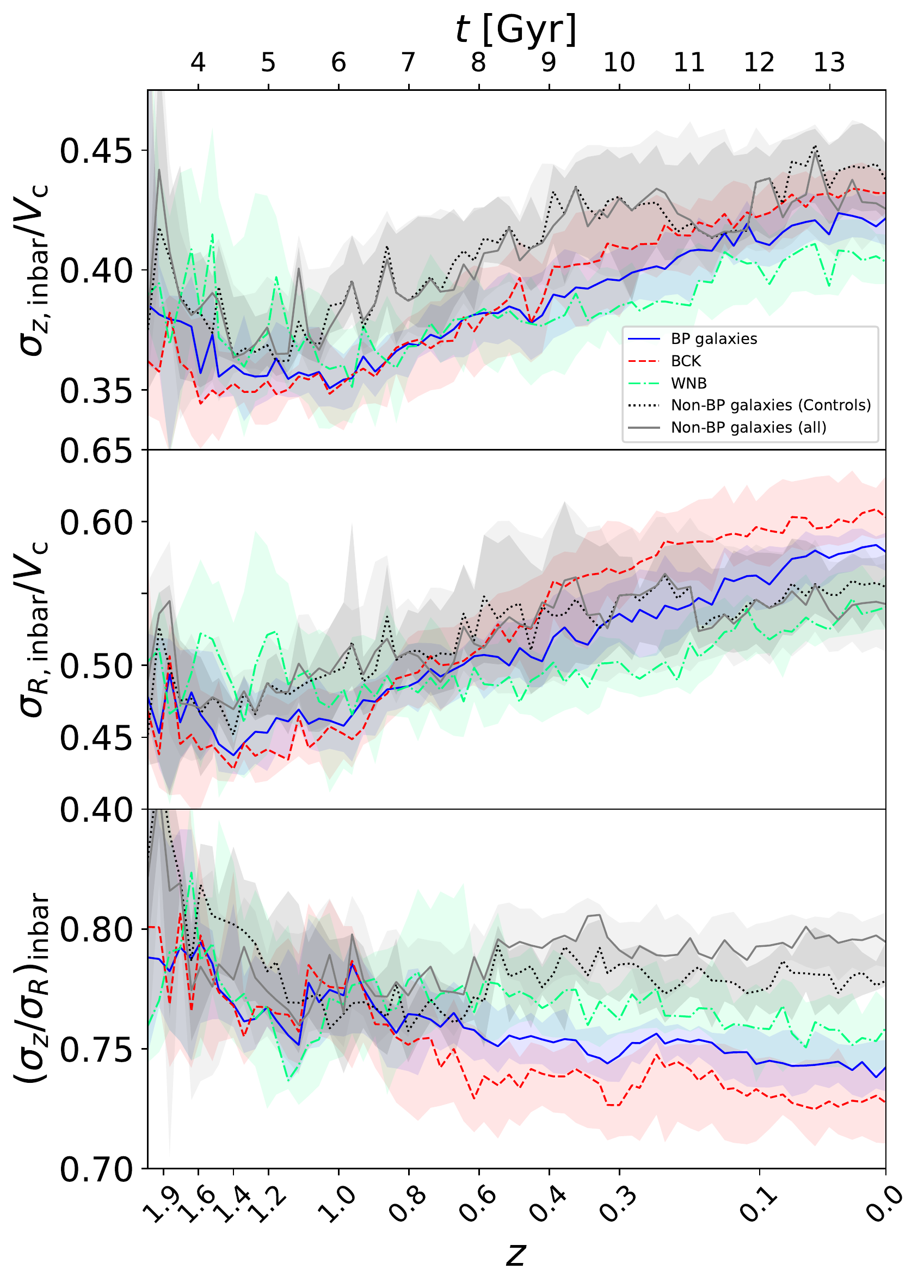}
  \caption{Evolution of the median $\sigzinbarvc$ (upper), $\sigRinbarvc$ (middle), and $\sigzsigRinbar$ (lower) ratios, for BP, BCK, WNB, non-BP bars and controls. \errortext}
  \label{fig:evolution_kinematics}
\end{figure}

We now compare the kinematics of the BP galaxies with those of the controls. 
Fig.~\ref{fig:evolution_kinematics} presents the evolution of velocity dispersions (normalised by the circular velocity at 2$\Reff$) $\sigzinbarvc$, $\sigRinbarvc$ and the ratio $\sigzsigRinbar$ (all within the bar). We focus on the bar region given our previous findings.
In the BP galaxies, both $\sigzinbarvc$ and $\sigRinbarvc$ rise steadily until the current epoch with $\sigRinbarvc$ perhaps rising faster. In the non-BP galaxies, $\sigRinbarvc$ flattens from $z\sim0.5$.

\begin{figure}
  \includegraphics[width=\hsize]{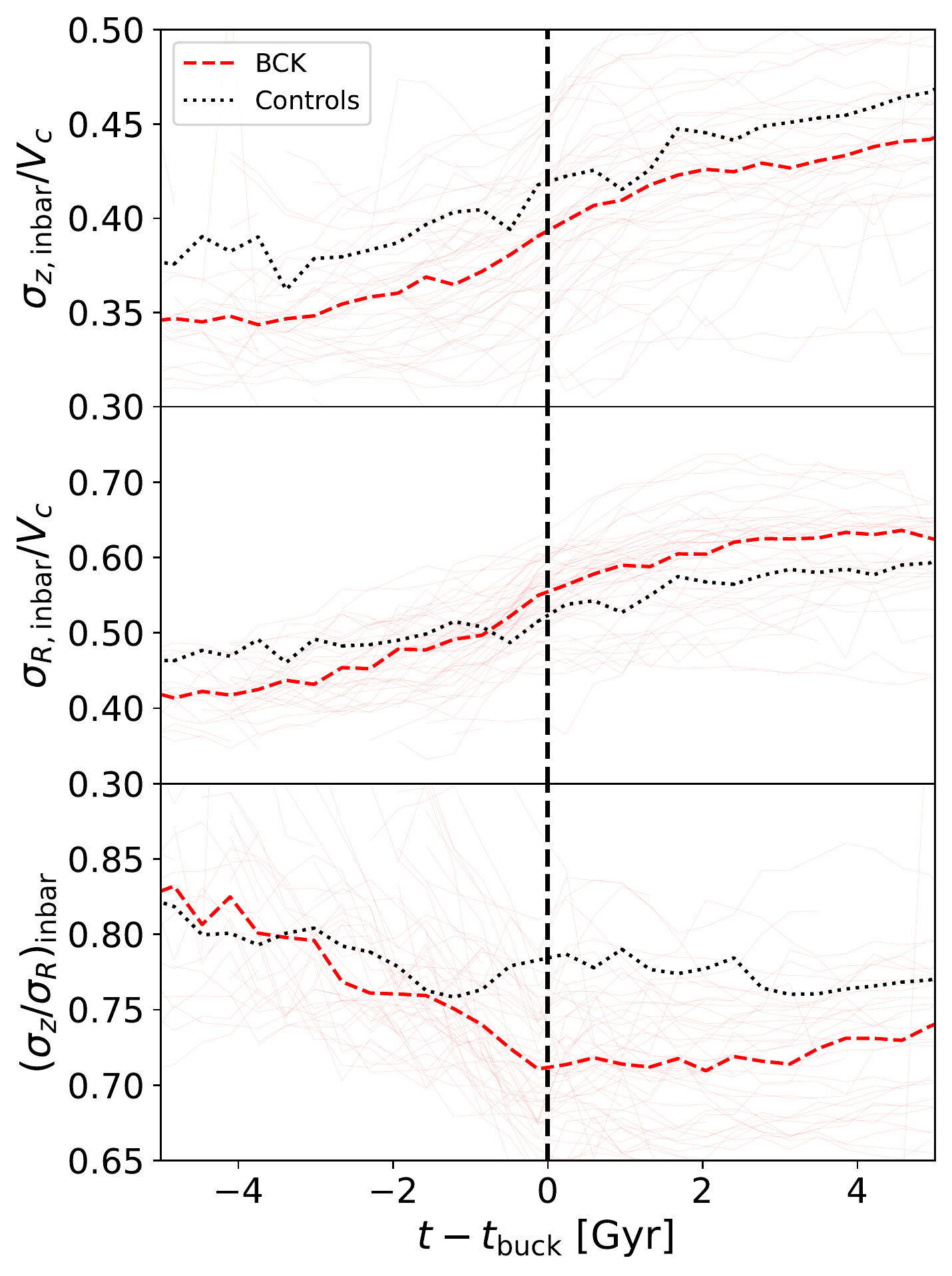}
  \caption{Evolution of the median ${\sigma_z}_\mathrm{,inbar}$ (upper), ${\sigma_R}_\mathrm{,inbar}$ (middle), $\sigzsigRinbar$ (lower) around the time of buckling $\tbuck$, for BCK (red dashed line) and control (black dotted line) galaxies. ${\sigma_z}_\mathrm{,inbar}$ and ${\sigma_R}_\mathrm{,inbar}$ are normalised by the circular velocity at 2$\Reff$. We show bins in $t-\tbuck$ containing 10 galaxies or more. We show individual BCK galaxies with thin red lines. The median $\sigzsigRinbar$ remains rather flat after buckling.}
  \label{fig:evolution_kinematics_sigzsigRinbar}
\end{figure}

The overall effect is seen in the ratio $\sigzsigRinbar$, which is key to the buckling instability, because it needs to reach low values (the bar becomes highly radially anisotropic) before buckling occurs \citep{toomre66, raha+91, sellwood96}. $\sigzsigRinbar$ declines steadily in the BP galaxies from $z\sim 1$, reaching a value of $\sim0.75$ at $z=0$. BP galaxies become more radially anisotropic as their bars strengthen, $\sigRinbarvc$ grows and $\sigzsigR$ declines. Eventually in the BCK sample, buckling is triggered \citep{sellwood96}. In the controls the decline is halted after $z\sim1$ (even increasing slightly). Their bars fail to strengthen significantly after this epoch, as seen in $\Atwomax$ (Fig.~\ref{fig:evolution_barstrength} upper left panel).

\begin{figure}
  \includegraphics[width=\hsize]{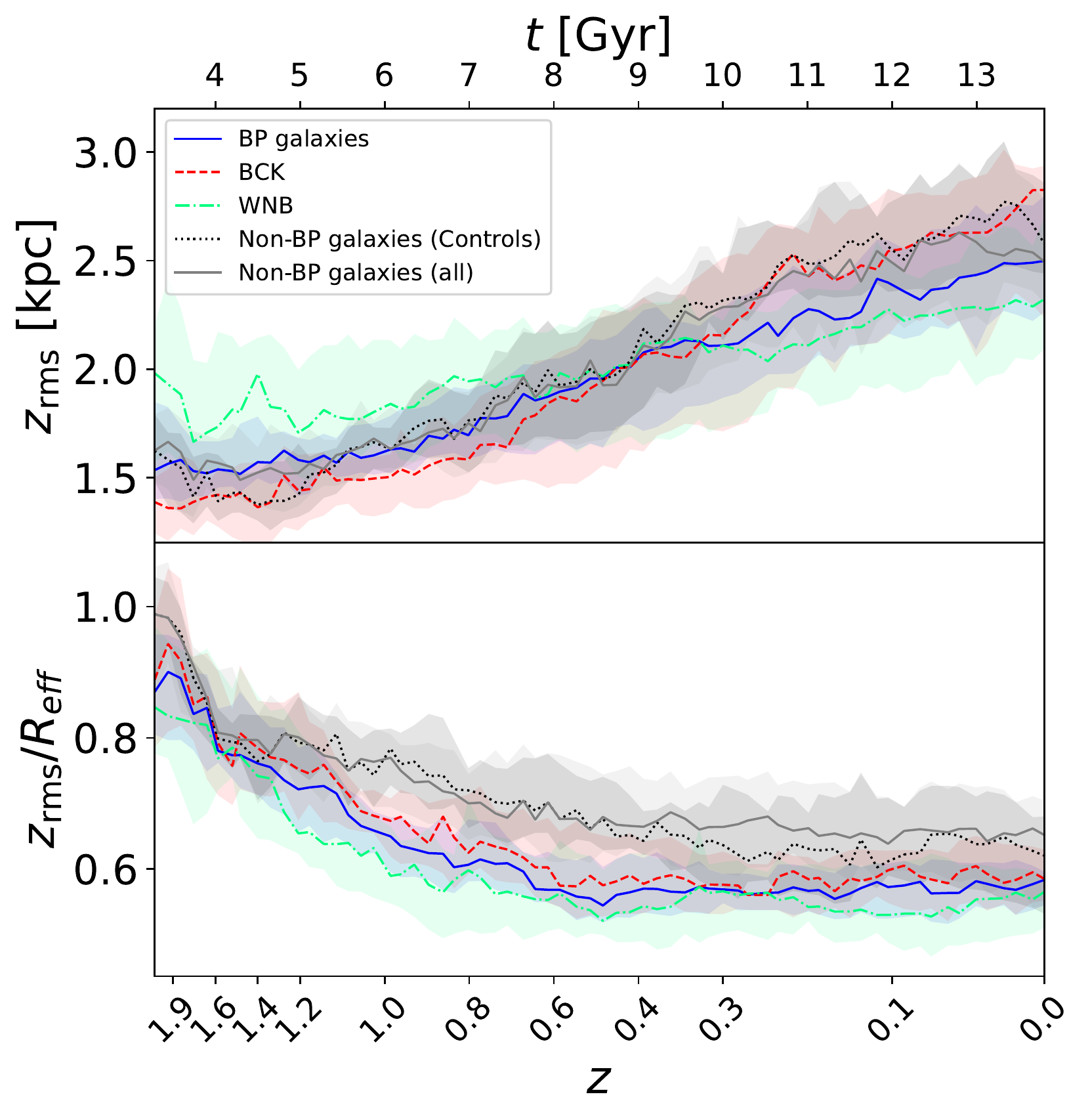}
  \caption{Evolution of the median $\zrms$ and $\zrms/\Reff$ for BP, BCK, WNB, control and non-BP galaxies. $\zrms$ is computed within $\Reff<R<2\Reff$, therefore mainly in the disc region. \errortext\ BPs occur in relatively somewhat thinner discs than non-BPs.}
  \label{fig:evolution_heights}
\end{figure}

In Fig.~\ref{fig:evolution_kinematics_sigzsigRinbar} we show the evolution of the velocity dispersions within the bar, but this time around $\tbuck$. The plot shows that, in contrast to isolated simulations \citep[e.g.][]{sotnikova_rodionov_2003,martinez-valpuesta_shlosman04,martinez-valpuesta+06,rodionov_sotnikova_13,lokas_2019,khoperskov_bertin_2017, li_shlosman+2023}, on average $\sigzsigRinbar$ remains rather flat after buckling. This implies that the ratio likely falls (or remains flat) for some galaxies after buckling. We have investigated this behaviour for a number of BCK galaxies, and found that the ratio was lower 2 Gyr after buckling in 22 of 52 BCK galaxies (the `fallers'). In almost all BCK galaxies, $\sigma_z$ rises after buckling then flattens, but for the fallers, $\sigma_R$ also rises through buckling and continues to do so, resulting in the ratio falling. In these galaxies, the bar strength continues to rise after buckling, somewhat unintuitively.

We explore whether this behaviour arises because some populations are immune to buckling by examining the evolution of the velocity dispersions in bins of time of formation, $t_\mathrm{form}$; we find large populations of stars within the fallers (and in those where the ratio remains flat) seemingly unaffected by buckling, as evidenced by the behaviour of $\sigma_z$ and $\sigma_R$ around $\tbuck$.
We compute the fraction of the stellar population which buckles in a simple way. For each group of stellar particles in bins of $t_\mathrm{form}$, we calculate the fractional change in $\sigma_z$ from $\tbuck - 0.5$ Gyr to $\tbuck + 1.5$ Gyr (to avoid noise around buckling itself), $f_{\Delta \sigma_z}$. If the $t_\mathrm{form}$ bin has $f_{\Delta \sigma_z} > 0.1$, we deem the population to have buckled. Hence for each galaxy we calculate the overall fraction of stars which have buckled ($f_\mathrm{buck}$). We find that the lower $f_\mathrm{buck}$, the smaller the rise in $\sigzsigRinbar$. Thus, our analysis suggests that the reason the median value of $\sigzsigRinbar$ remains flat after buckling is that there are a significant number of buckling galaxies which host a stellar population sufficiently hot to resist buckling.

Figure~\ref{fig:evolution_kinematics} also shows that non-BPs have higher $\sigzinbarvc$ or much of their evolution, hinting they may have been thicker (the control sample is controlled for stellar mass by design, so this is not a mass effect). Figure~\ref{fig:evolution_heights} (upper panel) shows the evolution of the mass-weighted root mean square height ($\zrms$). For this plot, we compute $\zrms$ and $\zrmsReff$ within $\Reff<R<2\Reff$, and so mainly in the disc region. On average, the BP galaxies have somewhat lower disc heights than the controls, especially after $z\sim0.4$, although the difference is not significant. Normalized by the effective radius (lower panel), the BP galaxies are thinner at $z=0$, and have been throughout their history. Controls and non-BP bars show similar evolution, so the trends found hold even when controlling for mass.

Thus, while BP galaxy discs have similar heights on average as non-BPs at $z=0$, normalised by their host galaxy size they have been thinner than non-BPs for most of their evolution down to $z=0$.

\subsection{Impact of mergers on the evolution of the samples}
\label{ss:impact_mergers}

\begin{figure}
  \includegraphics[width=\hsize]{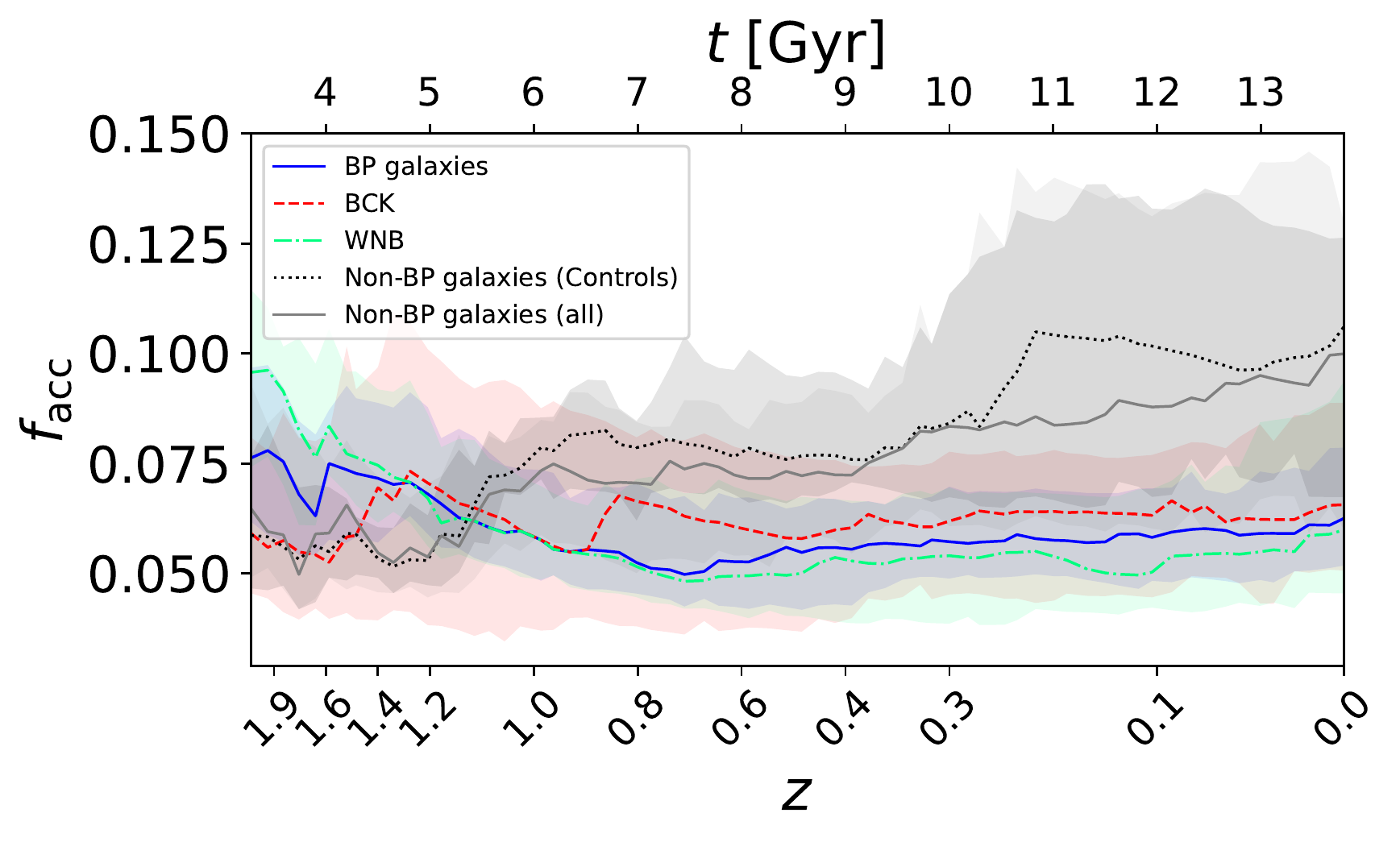}
  \caption{Evolution of the median accreted stellar mass fraction $f_\mathrm{acc}$ for BP, BCK, WNB, controls and non-BP barred galaxies. \errortext}
  \label{fig:evolution_mergers}
\end{figure}

Mergers can heat galaxies \citep[\eg][and references therein]{quinn+1993, reshetnikov_combes_1997_1, reshetnikov_combes_1997_2, sellwood+1998,purcell+2009,qu+2011} and weaken bars \citep[\eg][]{ghosh+21}. We explore the impact of merger and accretion events on the evolution of BPs.
Fig.~\ref{fig:evolution_mergers} shows the median accreted stellar mass fraction, $f_\mathrm{acc}$. The Figure reveals a divergence in behaviour between BP galaxies and non-BPs starting at $z\sim1$, after which the BP galaxies' accreted fraction remains constant. In contrast, for the controls, $\facc$ rises continually after this point (and their vertical heating is significantly higher throughout their history). Furthermore, the median time of the last major merger (mass ratio larger than $1/10$), is $z=0.89$ for the controls, but $z=1.41$ for the BPs, fully $1.85\Gyr$ earlier. Major merger activity for BPs ends much earlier than for the controls.

The epoch of this divergence in $\facc$ coincides with the growth in bar length and strength, and stall in the decrease in $\sigzsigR$ in the non-BPs (Sections~\ref{ss:bar_characteristics} and \ref{ss:kinematics_evol}). It also coincides with the time at which the disc's vertical and radial cooling stall in the controls (Fig.~\ref{fig:evolution_kinematics}).
We find $\facc=6.3$ per cent at $z=0$ for BP galaxies, but $10.0$ per cent for non-BP and control galaxies. Although the AIC$_\mathrm{c}$ for $\facc$ is higher than for bar length and strength (and so has a weaker discriminatory power as to whether a galaxy hosts a BP), these results imply that a sufficiently high accreted fraction suppresses BP formation by hindering the growth of a bar.

In TNG50, it may be that the heating of the disc prevents the controls' bars from strengthening, and that the accumulation of a satellite's mass in the central region of the host \citep[\eg][]{ghosh+21} weakens the bars after $z\sim1$. Mergers heat discs in all directions \citep{quinn+1993}, so the large exposure to mergers in the controls and non-BPs  holds $\sigzsigRinbar$ approximately constant on average, rather than declining as occurs in the BP bars.

We have checked Spearman rank correlations at $z=0$ and find a significant correlation between $\facc$ and $\zrms$ for both BP and non-BP samples ($\rho\sim0.8, p\lesssim10^{-23}$), and between $\facc$ and $\Reff$ ($\rho\sim0.7, p\lesssim10^{-15}$). These results support the notion that merger activity influences a galaxy both vertically and radially, but that height may be impacted more, resulting in control and non-BP galaxies' higher relative heights. We also find a negative correlation between $\facc$ and $\Atwomax$, which we discuss in Section~\ref{s:discussion}.

We also explore whether environment plays any role in BP formation. Of the 608 disc galaxies, 413 (68\%) are centrals and 195 (32\%) are satellites. Of the 191 barred galaxies, 130 (68\%) are centrals and 61 (32\%) are satellites. Of the 106 BP galaxies amongst those bars, 77 (73\%) are centrals and 29 (27\%) are satellites. These $z=0$ fractions are consistent, hinting that a galaxy's classification as a central or satellite is not relevant to whether it hosts a BP.

\begin{figure}
  \includegraphics[width=\hsize]{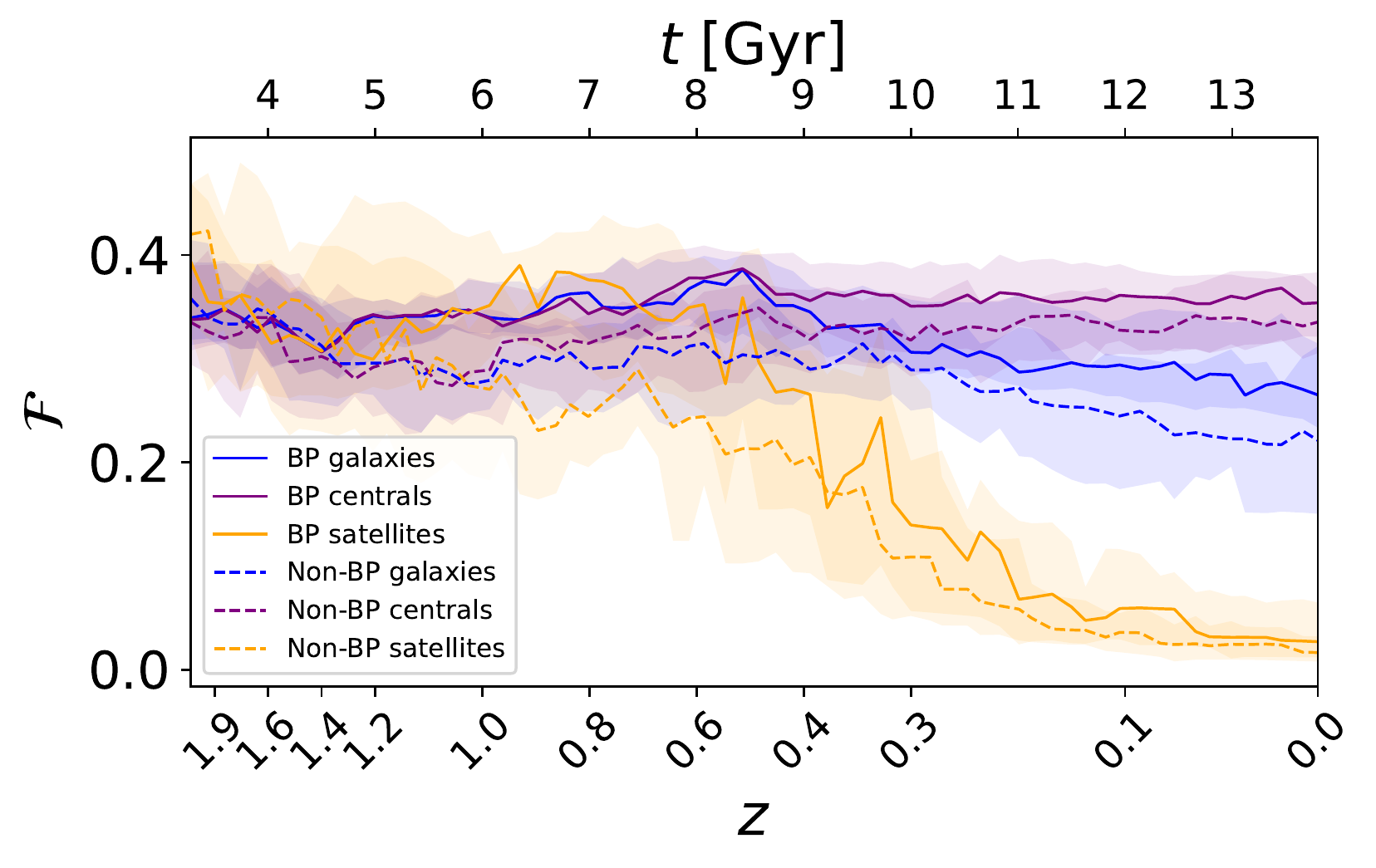}
  \caption{Evolution of the median $\mathcal{F}=M_{\rm{gal}}(z)/M_{\rm{group}}(z)$ for BPs and non-BPs. \errortext\ BPs and non-BPs (whether satellites or centrals) share approximately the same $\mathcal{F}$ for most of their evolution.}
  \label{fig:cent_sat_galmass_totalmass}
\end{figure}

We compute $\mathcal{F}=M_{\rm{gal}}(z)/M_{\rm{group}}(z)$, where $M_{\rm{gal}}(z)$ is the total mass of the galaxy (dark matter +  gas + stars) within $10\Reff$, and $M_{\rm{group}}(z)$ is the total mass (dark matter +  gas + stars) of the friends of friends group in which the galaxy is located (supplied as part of the IllustrisTNG public data release), at each $z$. We expect galaxies with small $\mathcal{F}$ to be more influenced by tidal effects than galaxies with large $\mathcal{F}$. If environment plays a major role in determining whether a galaxy forms a BP, we expect a difference in $\mathcal{F}$ between BPs and non-BPs. We show the evolution of median $\mathcal{F}$ in Fig.~\ref{fig:cent_sat_galmass_totalmass}. Although $\mathcal{F}$ is systematically somewhat lower for non-BPs (dashed lines) than BPs (solid lines), it is not statistically significant except perhaps for $z>0.5$ for satellites. We conclude that whether a galaxy is a satellite or central, the BP and non-BP galaxies share \textit{approximately} the same $\mathcal{F}$, but with controls having somewhat lower values on average.

The AIC$_{\mathrm{c}}$ for $\mathcal{F}$, is 266.48, 13.9 higher than for $\facc$ (Table~\ref{tab:logistic_regression}). So environment matters only indirectly, via its legacy in $\facc$ in determining the probability of hosting a BP. We conclude that merger activity as measured by $\facc$ is a significant contributor to non-BP galaxies' relatively thicker discs and weaker bars, and the suppression of BP formation.

\begin{figure}
  \includegraphics[width=\hsize]{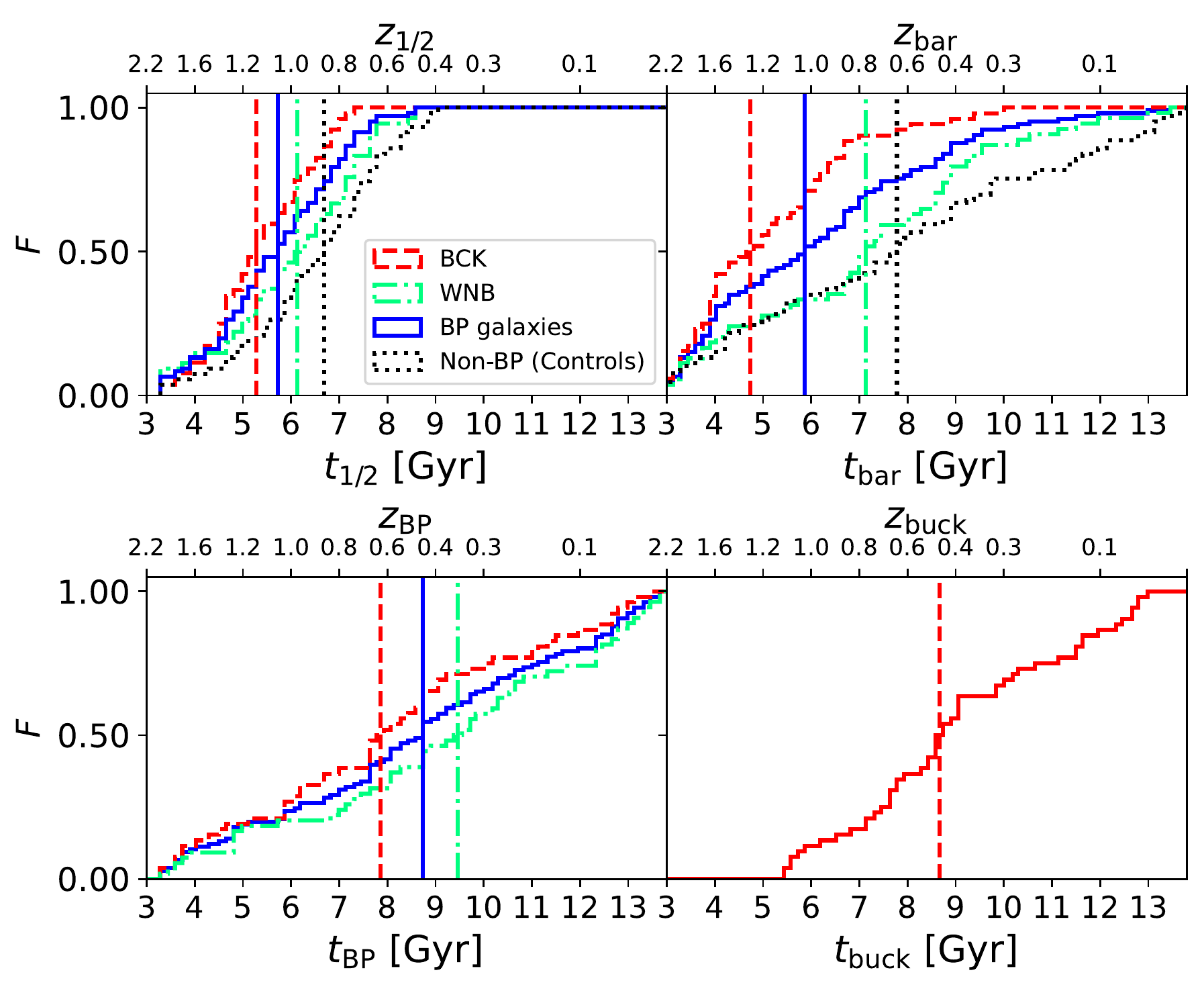}
  \caption{Cumulative distributions of the times by which the stellar mass of the samples reaches half its value at $z=0$ ($t_{1/2}$) (top left panel), $\tbar$ (top right panel),  $\tbp$ (lower left panel) and $\tbuck$ (lower right panel). Medians are shown as vertical dashed lines. Bars in BP galaxies form much earlier than those of the same stellar mass without BPs.}
  \label{fig:event_times}
\end{figure}

\section{Downsizing and bar/BP age}
\label{s:downsizing}

Fig.~\ref{fig:event_times} shows the cumulative distribution of the time at which each galaxy assembles half of its $z=0$ stellar mass (\thf). On average, BP galaxies assemble their stellar mass $\sim1\Gyr$ earlier than the controls. The figure also shows the times of bar and BP formation (\tbar\ and \tbp, respectively) and times of buckling. It shows that bars in the BP galaxies form $\sim2\Gyr$ earlier on average than those in the controls. Bars in BCK galaxies form $\sim2 \Gyr$ earlier than those in WNBs, even though there is no statistically significant difference in their mass distribution. Thus BP bars are on average older than non-BP bars, and bars in strongly buckled galaxies are on average older than in WNB and non-BP barred galaxies. Thus a `sequence' of bar age, young to old (controls$\;\xrightarrow{}$WNB$\;\xrightarrow{}$BCK), exists. Clearly galaxy mass assembly and bar formation timescales influence BP formation. This motivates us to further compare old and young bars. We note in passing that the bottom panels of Fig.~\ref{fig:event_times} show that the BP formation and buckling rates are consistent with being constant with time, albeit with the caveat that we only examine those BPs extant at $z=0$. Furthermore, the lower left panel showing $\tbp$ is in reasonable agreement with \citet{kruk+19}, who estimated the earliest BPs to form at $z\sim1$ (note that their work examined BPs at those earlier redshifts, whereas this study examines only $z=0$ BPs).  

\begin{figure*}
  \includegraphics[width=\hsize]{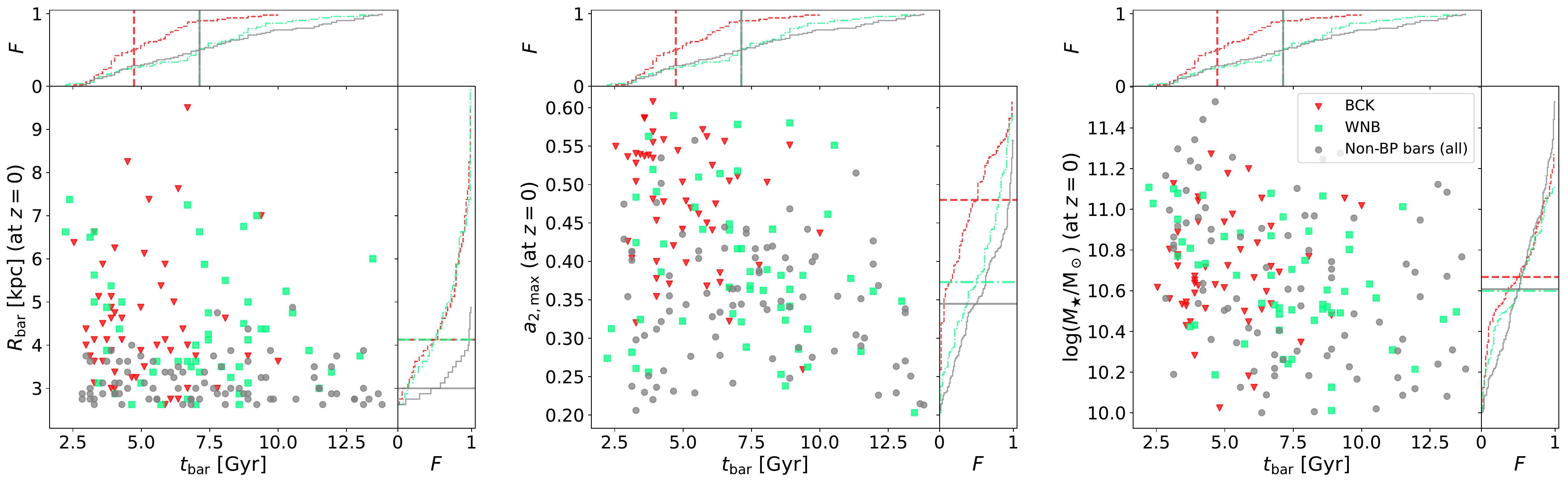}
  \caption{The distribution of barred galaxy samples in the  ($\tbar$,$\Rbar$), ($\tbar$,$\Atwomax$)- and ($\tbar$,$\logMstar$)- planes, identifying BCK, WNB and non-BP galaxies. The side panels represent cumulative distributions of the parameter on the respective axis, with their median shown in vertical (top) and horizontal (right) dashed lines. For clarity, the $y$-axis limit in the top panel is set to 10 kpc (there is one WNB galaxy at $\Rbar=13.6$ kpc, $\tbar=9.55$ Gyr).}
  \label{fig:t_bar_by_mass}
\end{figure*}

Fig.~\ref{fig:t_bar_by_mass} shows individual galaxies as points on three 2D planes at $z=0$. The left panel (the ($\tbar,\Rbar$)-plane) shows that for the non-BP galaxies, bars are relatively short irrespective of their age. For BP galaxies, in contrast, older BP bars tend to be longer. The panel also shows that at a given bar age, BPs are hosted in galaxies with larger bars. The middle panel of Fig.~\ref{fig:t_bar_by_mass} (the ($\tbar$, $\Atwomax$)-plane) shows that older bars (lower $\tbar$) are stronger than younger bars, contributing to the relative lack of BPs amongst younger bars.

\begin{figure}
  \includegraphics[width=\hsize]{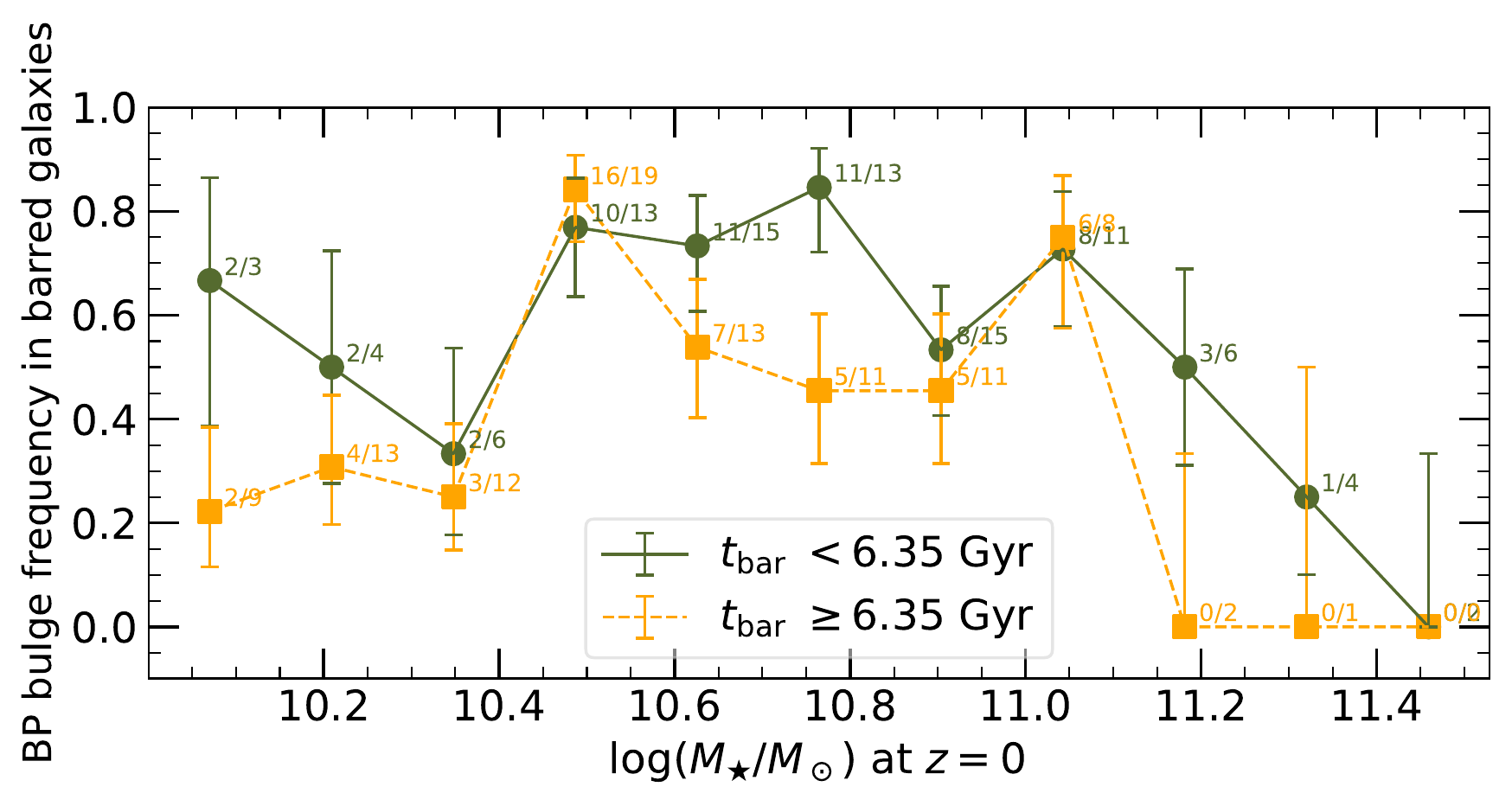}
  \caption{The fraction of barred galaxies hosting BPs at $z=0$, split into young (orange dashed line, squares) and old (olive solid line, circles) populations, defined by median $\tbar$. The denominator in the fraction within each group is the number of barred galaxies within that group. Error bars are the $68$ per cent (1$\sigma$) confidence limits from the \citet{wilson_1927} binomial confidence interval.}
  \label{fig:frac_BP_tbar_split}
\end{figure}

The right panel shows the ($\logMstar$,$\tbar$)-plane and reveals that the bars in high-mass galaxies formed earlier than those in low-mass galaxies. A similar trend was found in TNG100 by \citet{rosas_guevara+2020}. This `downsizing' trend has been noted in observations by \citet{sheth+08} and more recently by \citet{cuomo_etal_2020}. We have checked the correlation between $\tbar$ and $\thf$, finding them to be monotonically related\footnote{Spearman $\rho=0.52$ with $p$-value = $1.4\times10^{-14}$.}, albeit with some scatter. Therefore, this result seems to be driven by stellar downsizing, where the stars in more massive galaxies tend to have formed earlier \citep{neistein+2006, pilyugin_thuan_2011}. At lower mass ($\logMstar \lesssim 10.3$) there are fewer BP galaxies compared to non-BP galaxies than at higher mass (Fig.~\ref{fig:z0BPfracdist}). This is driven by the lack of \emph{old} bars (low $\tbar$) at low mass, particularly BCK galaxies (a much weaker effect is also seen for WNB galaxies) -- see the CDF at the top. This is confirmed in Fig.~\ref{fig:frac_BP_tbar_split}, which shows $\fbp(\Mstar)$ split by the median $\tbar$ for all barred galaxies (6.82\Gyr). Younger bars are less likely than older bars to host BPs.

So even though overall $\tbar$ has a weaker discriminatory power as to whether a galaxy hosts a BP than $\Rbar$ (Table \ref{tab:logistic_regression_single_variable}), none the less bar age plays a role in the emergence of a BP.

\begin{figure*}
  \includegraphics[width=\hsize]{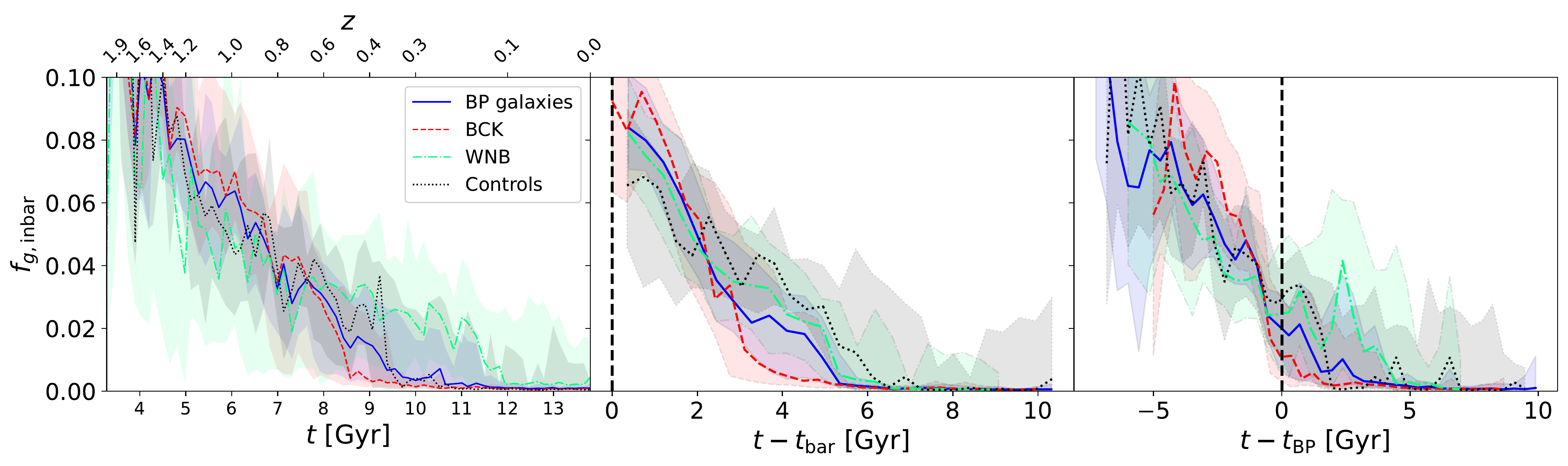}
  \caption{Evolution of the median $\fginbar$ for WNB, BCK and BP galaxies, and for the controls. Left panel: the evolution of the median $\fginbar$ from $z=2$. Middle panel: the evolution of the median $\fginbar$ as a function of $\Delta t= t-\tbar$ for each galaxy. $\fginbar$ is defined only after $t-\tbar=0$ Gyr, represented by the vertical dashed black line. Right panel: the evolution of the median $\fginbar$ as a function of $\Delta t= t-\tbp$ for each galaxy (each control galaxy, with no BP by definition, is assigned the same $\tbp$ as its mass-matched BP galaxy). The vertical dashed black line represents the time of BP formation, $t=\tbp$. In the two right panels we show only bins in $\Delta t$ containing 10 or more galaxies. In the middle panel, we exclude 14 barred galaxies whose bars form before $z=2$ since at these earlier times, the bar radius has high uncertainty. \errortext}
  \label{fig:gas}
\end{figure*}

\section{The role of gas}
\label{s:gas}

Simulations of isolated galaxies show that high gas content suppresses buckling \citep{berentzen+98,debattista+06,berentzen+07,wozniak_michel-dansac09,lokas_2020}, whereas \citetalias{erwin_debattista17} found no observational link between the presence of BP bulges and the global gas content (as measured by $M_{\rm H{I}}/M_\star$), after controlling for the dependence of gas fraction on stellar mass. These seemingly contradictory results motivate us to test for any role for gas in TNG50 BP bulge formation.

In the studies cited above, the gas fractions required to suppress buckling vary substantially\footnote{Some of these studies define $f_g$ as a fraction of the disc mass, which is different from our definition. See Table~\ref{tab:symbols}.}. Amongst isolated simulations, \citet{berentzen+07} found buckling was suppressed for $f_g \gtrsim 0.03$, \citet{wozniak_michel-dansac09} found buckling suppressed for $f_g\sim0.1$, and \citet{lokas_2020} found that $f_g = 0.3$ was sufficient, with buckling occurring for $f_g=0.2$. In TNG50 we find median $f_g\sim0.4$ at $\tbp$, the fraction in the BCK sample being somewhat lower than that in the WNB and control samples.

Tables \ref{tab:logistic_regression_single_variable} and \ref{tab:logistic_regression_multi_variable} show that neither $f_g$ nor $\fginbar$ are able to discriminate between BPs and non-BPs significantly compared to bar length and strength (AIC$_\mathrm{c}$ for $\fginbar$ is fully 72.0 higher than for $\Rbar$). Fig.~\ref{fig:gas} shows the evolution of $\fginbar$, also as a function of $\Delta t_\mathrm{bar} = t-\tbar$ and $\Delta t_\mathrm{BP} = t-\tbp$. In the left panel, we see no significant evolutionary differences between BP galaxies and controls.

At $\tbar$, $\fginbar \sim0.07$ for BPs and controls. Since the difference at $\tbar$ is insignificant, galaxies destined to host BPs are not sensitive to the gas fraction at the time their bars form. Likewise, at $\tbp$, $\fginbar$ is similar in BP galaxies and in the controls ($\sim0.02$). Two-sample KS tests confirm the lack of significant differences in the distributions of $\fginbar$ at $\tbar$ and $\tbp$ for BPs and controls ($p\sim0.06$ and $0.02$, respectively). It seems that the gas fraction within the bar does not significantly inhibit the formation of BPs (either BCK or WNB). Qualitatively similar findings apply for the global gas fraction, in agreement with \citetalias{erwin_debattista17}.

We conclude that the gas fraction within the bar is so low, that in TNG50 it is not responsible for whether a BP forms or not.

\section{Discussion}
\label{s:discussion}

\subsection{BP fraction at low and high stellar mass}
\label{ss:BP_frac_discussion}

We now interpret the behaviour of $\fbp(\Mstar)$ seen in Fig.~\ref{fig:z0BPfracdist}. We assign an approximate stellar mass at which the plateau in $\fbp(\Mstar)$ begins, using the logistic regression fit for `All BPs' in Table~\ref{tab:logistic_regression}. We set this at the point where $\fbp(\Mstar)$ reaches 95\% of the asymptotic fraction $K$ $(=0.59)$, which yields $\logMstar=10.50$. We are especially interested in why there are so few BPs at low stellar mass, since they mirror the observations. We also interpret why the profile has a plateau, why this occurs at $\logMstar\gtrsim10.5$, and why it asymptotes at $\sim 0.6$ in TNG50, whereas in observations the fraction saturates at $1$ at high mass (\citetalias{erwin_debattista17} and EDA23).

\begin{figure}
  \includegraphics[width=\hsize]{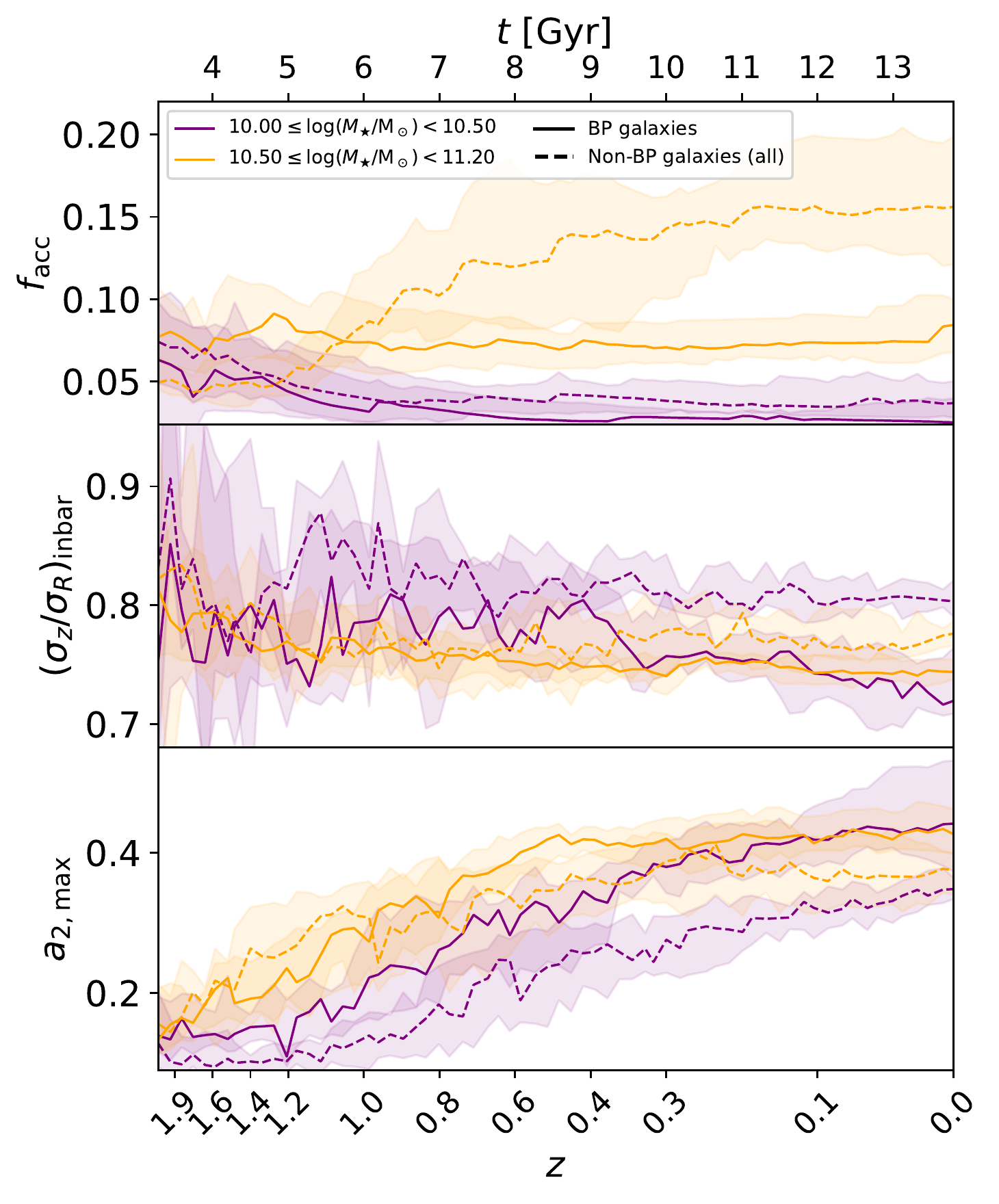}
  \caption{Evolution of the median $\facc$ (upper panel), $\sigzsigRinbar$ (middle panel) and $\Atwomax$ (lower panel) for BP (solid lines) and non-BP galaxies (dashed lines) by bins in $\Mstar$ at $z=0$, selected to match two key regions in the $\fbp(\Mstar)$ distribution (Fig.~\ref{fig:z0BPfracdist}). The lowest and highest mass bins contain 28 and 78 galaxies in the BP sample, respectively. The corresponding numbers in the non-BP barred sample are 34 and 46. \errortext}
  \label{fig:evol_by_mass_bin}
\end{figure}

We define two mass bins: a low-mass bin before the plateau ($10.0\le\logMstar<10.5$), and a high-mass bin ($10.5\leq\logMstar\leq11.2$), the latter covering the plateau region. This binning strategy is at the expense of having different numbers of galaxies in each bin (28 and 78 BP galaxies in the respective mass bins, and 34 and 42 galaxies for the non-BP galaxies). There are just 10 (of $\numbars$) galaxies with $\logMstar>11.2$, and we do not show this bin on the plots for clarity since the dispersion in this bin would dominate the figures. Figure~\ref{fig:evol_by_mass_bin} shows the evolution of various properties in each mass bin.

We examine first the low-mass regime where the BP fraction is low ($\lesssim 0.3$) but growing to $\sim0.6$ at its high-mass end (Fig.~\ref{fig:z0BPfracdist}). In Fig.~\ref{fig:evol_by_mass_bin}, the low-mass bin (purple lines) has much lower $\facc$ than the higher mass bin. Despite this quiescent merger history, $\fbp$ is low at this mass. Therefore, we can eliminate merger activity as the cause of suppression of the BP fraction at low mass.

\begin{figure}
  \includegraphics[width=\hsize]{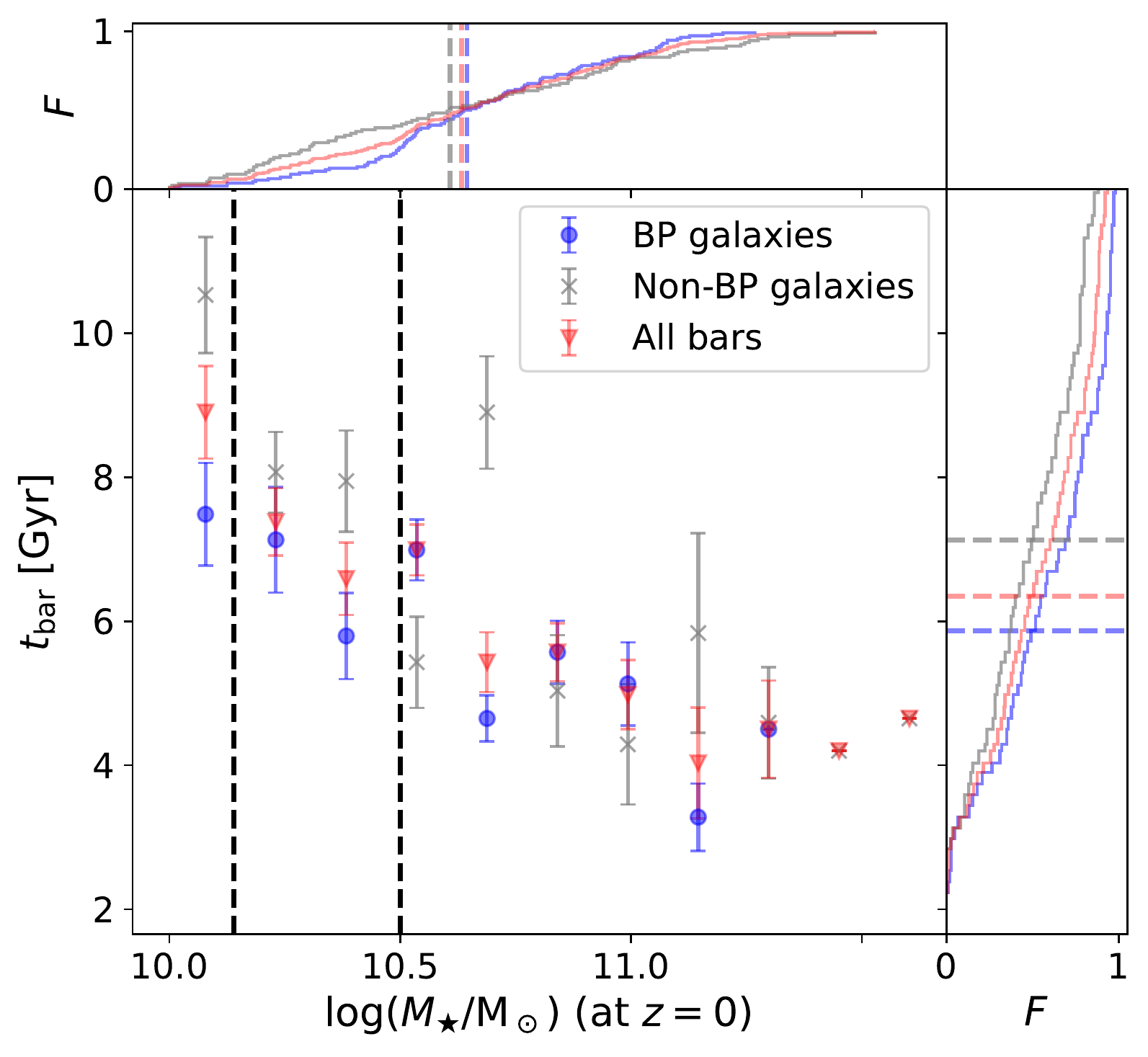}
  \caption{Median $\tbar$ in bins of stellar mass for BP (blue circles), non-BP (grey crosses) and all barred galaxies (red triangles). The error bars represent the standard error on the median. The side panels represent cumulative distributions of the parameter on the respective axis, with their median shown in vertical (top) and horizontal (right) lines. Two vertical dashed black lines mark the turning point and the beginning of the plateau region in $\fbp(M_\bigstar)$ ($\logMstar=10.13$ and $10.50$, respectively). At low mass, bars are significantly younger (higher $\tbar$) than at higher mass, contributing to suppression of the BP fraction.}
  \label{fig:median_t_bar_by_mass}
\end{figure}

We have shown that young bars are weaker than older bars, resulting in fewer BPs. Figure~\ref{fig:median_t_bar_by_mass} shows the median $\tbar$ by stellar mass bin. It clearly shows the dominance of younger bars (higher $\tbar$) in the low-mass bin. At lower mass, BP bars are older (lower $\tbar$) than non-BP bars. Furthermore, amongst the BPs, we find that at low mass, there are more WNB than BCK galaxies (Fig.~\ref{fig:z0BPfracdist}).  Fig.~\ref{fig:evol_by_mass_bin} shows that $\sigzsigRinbar$ is higher at low mass which suppresses buckling, since buckling requires low $\sigzsigRinbar$. Also, Fig.~\ref{fig:BP_frac_by_bar_length_and_strength} shows that at low mass, few bars are long, and those that are tend to have a high BP fraction. So in TNG50, at low mass, it is bar youth and their shorter, weaker bars, which cause a low BP fraction.

\begin{figure*}
  \includegraphics[width=\hsize]{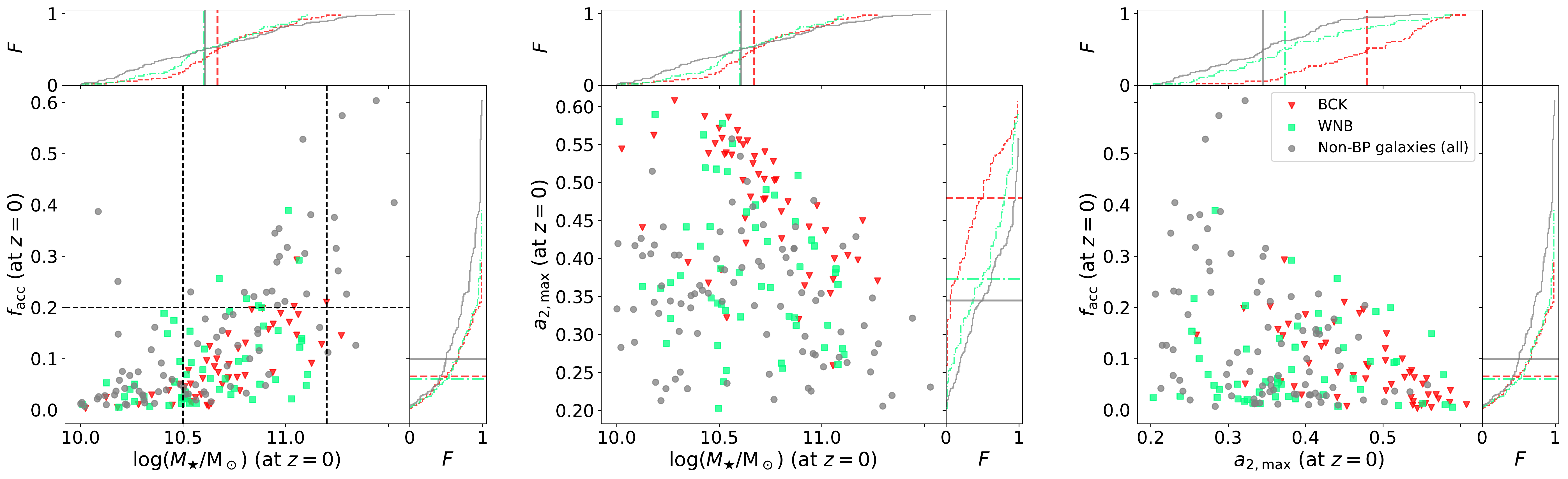}
  \caption{The distribution of barred galaxy samples in the ($\logMstar, \facc$) (left panel), ($\logMstar, \Atwomax$) (middle panel) and ($\Atwomax$, $\facc$) (right panel) planes at $z=0$, identifying BCK, WNB and non-BP galaxies. The side panels represent cumulative distributions of the parameter on the respective axis, with their median shown in vertical (top) and horizontal (right) lines. Dashed lines are inserted at $\facc=0.2$ and $\logMstar=10.5$ \& $11.2$ in the left panel to guide the eye.}
  \label{fig:facc_a2max_Rbar_by_mass}
\end{figure*}

We now examine the high-mass regime. For all high-mass galaxies $\facc$ is higher from $z\sim1$ than for low-mass galaxies (Fig.~\ref{fig:evol_by_mass_bin}, upper panel). This is consistent with recent work by \citet{guzman-ortega+2023} who found that the merger fraction increases with stellar mass in both TNG50 and observations. Figure~\ref{fig:facc_a2max_Rbar_by_mass} shows the ($\logMstar,\facc$), ($\logMstar,\Atwomax$) and ($\Atwomax,\facc$)~planes for the BP and non-BP galaxies. It demonstrates that few BP galaxies with $\facc\ge0.2$ have $\logMstar\gtrsim10.5$, where the plateau in $\fbp(\Mstar)$ begins. The figure also shows (middle panel) that, on average, bars are strongest at $\logMstar\sim10.5$, but are weaker for higher mass galaxies. The left panel of the figure suggests this is due to a higher accreted fraction. The right panel reinforces this conclusion by showing the relation between $\facc$ and $\Atwomax$, which reveals a declining bar strength with increasing $\facc$ at $z=0$ (Spearman $\rho=-0.313$, $p=1.1\times10^{-5}$). This is strong evidence that a greater accreted fraction is associated with weaker bars. Since $\facc$ increases with mass, why then does $\fbp$ remain approximately flat at higher mass instead of declining?

We showed in Section~\ref{ss:bar_characteristics} that in the barred sample, there is no monotonic relation between $\logMstar$ and $\Rbar$, so the increase in $\facc$ is not compensated for by an increase in bar length. Figure~\ref{fig:evol_by_mass_bin} shows a large difference in the high-mass bin between $\facc$ in the BPs (median $8$ per cent) and the non-BPs (median $17$ per cent), with the former having a noticeably narrower dispersion. This hints that a BP can develop up to a certain level of accretion (and consequent heating), but no further. Thus the formation of a BP may be neutral to the increase in merger activity as mass increases up to a point, holding $\fbp$ flat. However, for the few galaxies with $\logMstar>11.2$, perhaps the heating is so great that BP formation is mostly suppressed, which might explain why their $\fbp$ is so low ($\lesssim0.1$). The overall numbers at high mass are so small however, that we cannot draw any firm conclusions.

\begin{figure}
  \includegraphics[width=\hsize]{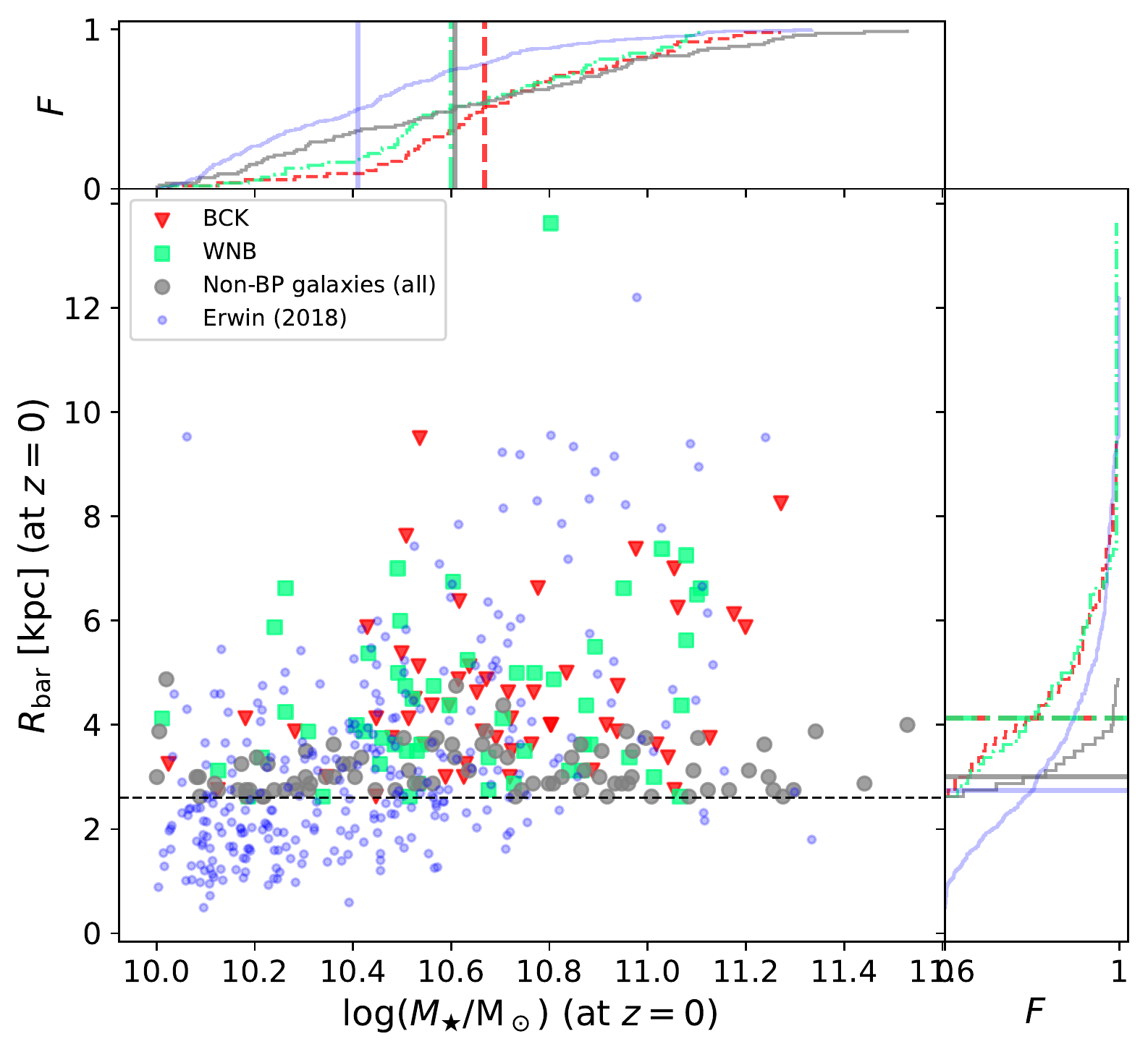}
  \caption{The distribution of barred galaxy samples in the ($\logMstar, \Rbar$)-plane at $z=0$, identifying BCK, WNB and non-BP galaxies. The side panels represent cumulative distributions of the parameter on the respective axis, with their median shown in vertical (top) and horizontal (right) panels. We overlay in light blue circles observational data from \citet{Erwin_2018} (see Section~\ref{ss:buckling_implications}), with a cut on $\logMstar\ge10.0$ to match the TNG50 sample (but without a cut on $\Rbar\ge2.6$ kpc). For these data, $\Rbar$ is the deprojected semi-major axis length of the bar. For the \citep{Erwin_2018} data, to match the TNG50 sample, we show cumulative lines only for $\Rbar\ge2.6$ kpc. The horizontal black dashed line represents $\Rbar=2.6$ kpc.}
  \label{fig:Mstar_versus_Rbar}
\end{figure}

In summary, our analysis shows that, in TNG50, the reason for the non-BP bars' weakness is different in each mass regime -- bar youth at low-mass, and merger activity at high mass. Downsizing (in the sense of bars being younger, hence weaker, in lower mass galaxies and older in higher mass galaxies) governs the \emph{shape} of $\fbp(\Mstar)$ (the fact that it increases with mass to a plateau). Only above a certain mass are there enough sufficiently dynamically `mature' bars to form BPs in significant numbers. The \emph{value} of $\fbp(\Mstar)$ in a given mass bin is dependent both on the proportion of young and old bars, and on the accretion history of the galaxies in that mass bin. More generally, the heating of a barred galaxy is a key factor in determining its susceptibility to BP formation.

\subsection{Comparison with observations}
\label{ss:compare_obs}

We caution that there are differences in the way barred galaxies and BPs have been selected here, and in observational studies. None the less, a comparison can provide further insights into how BPs develop in nature.

Owing to the caveats outlined in Section~\ref{s:TNG50_overview} we consider whether the BPs identified in TNG50 can be detected observationally. We sampled six approximately MW-stellar-mass galaxies from the BP sample, half of which were in the BCK subgroup, the other half in WNB. We generated mass maps with the galaxies at a few of the preferred orientations as determined by \citetalias{erwin_debattista17}. In only one case did we detect the `box+spurs' morphology of a BP \citep{erwin_debattista13}. We believe our BP detection method therefore is able to find weak BPs which would be hard to detect observationally.

TNG50 does not match the stellar mass distribution of the BP fraction seen in observations (Fig.~\ref{fig:z0BPfracdist}), although the overall fraction of 55\% is close to, but somewhat lower than, observations. \citet{yoshino_yamauchi15} found a BP fraction of $\sim66$ per cent, and \citet{kruk+19} found $\sim70$ per cent for example. It does however match the overall profile \emph{shape} for $10.0\lesssim\logMstar\lesssim11.2$ (we have too few galaxies above this mass to make a credible assessment). Furthermore, this study finds a `turning point' where $\fbp$ reaches half its asymptotic high-mass value at $\logMT\sim 10.1$, matching the sample of EDA23. \citetalias{erwin_debattista17} found an upturn at $\logMT\sim 10.4$ in their full sample ($\ie$ without the bar radius cut we adopt). In a sample of 1925 edge-on BP galaxies \citet{marchuk_etal_2022} also found an upturn in $\fbp$ at $\logMT\sim 10.4$, albeit with a sample of unknown bar radius distribution. Thus the turning point we find is similar to that found in observations.

At low mass, where TNG50 closely matches the BP fraction in observations, bar youth (and hence weakness) may be a key driver of low BP fraction in nature. At higher mass it may be that the heating from mergers in the real Universe is simply not as efficient at inhibiting BP formation as we find in TNG50. At intermediate to high mass, galaxies in TNG50 may be too kinematically hot. Indeed, \citet{wang_l_etal_2020} found that at $z=0.15$, the major merger fraction in TNG100 matches observations reasonably well at $10.3\lesssim\logMstar\lesssim10.7$ (close to the turning point we find in $\fbp$), but is considerably ($\sim50$ per cent) higher than observations above this mass (see their Fig.~10). Observations show that bar length and strength are secondary variables in explaining whether a galaxy has a BP or not \citepalias[EDA23 and][]{erwin_debattista17}. Had the excessive heating been absent in TNG50, we speculate that $\fbp$ would in fact plateau at a higher level, and that $\logMstar$ would have the highest discriminatory power in determining whether a galaxy had a BP or not, rather than bar characteristics (Tables~\ref{tab:logistic_regression_single_variable} and \ref{tab:logistic_regression_multi_variable}).

Despite the excessive heating overall, we have found that TNG50 BP galaxies have had a more quiescent merger history and had (relative to $\Reff$) thinner discs than non-BP galaxies at $z=0$ (Sections~\ref{ss:kinematics_evol} and \ref{ss:impact_mergers}).

\subsection{Implications for the frequency of buckling}
\label{ss:buckling_implications}

We have found no less than $\sim50$ per cent of BP bulges at $z=0$ in TNG50 which had at least one strong buckling episode in their history. There is reason to believe that, if anything, this fraction is an underestimate of the fraction in the real Universe.

\citet{merritt_sellwood94} showed that buckling can occur provided that a significant population of bar stars have vertical  frequencies, $\nu$, which are larger than the frequency with which they encounter the vertical forcing from the bar. For an $m=2$ bending instability, this vertical forcing has frequency $2(\Omega-\Omega_p)$, where $\Omega$ is the angular frequency of the star and $\Omega_p$ is the pattern speed of the bar. Thus anything that causes $\nu$ to decrease inhibits buckling.

In TNG50, $\nu$ may be small for many stars. The softening length of TNG50 is $288\pc$ at $z=0$. This is rather large in comparison to the thickness of the average disc galaxy (for example the Milky Way's thin disc has scale height $300 \pc$, \citealt{bland-hawthorn_gerhard16}). As a result, the vertical frequency of stars near the centres of galaxies are lower than they would be in a real galaxy with the same vertical density distribution (since the restoring force towards the plane is lower). \citet{merritt_sellwood94} show that, contrary to what might be naively expected, larger force softening {\it suppresses} buckling, rather than enhances it, because it reduces $\nu$.

Furthermore, \citet{ludlow_etal_2021} find that the relatively low dark matter (DM) particle resolution in cosmological simulations results in collisional heating both radially and vertically, albeit with a relatively small effect for TNG50 (the authors estimate a consequent increase in vertical scale height of $0.18 \kpc$). In their recent work, \citet{Wilkinson+2023} showed that simulations are especially susceptible to spurious collisional heating if the number of DM particles $\lesssim 10^6$. In our work the median $\log(n_\mathrm{dark})\sim 5.7$ (within $10R_\mathrm{eff}$) for the barred sample at $z=0$ in our sample and 90\% have $\log(n_\mathrm{dark}) < 6.0$. So collisional heating may increase the thickness of the galaxies somewhat, especially for the lower-mass galaxies where the number of DM particles is small. This naturally leads to fewer BPs than if the galaxies were resolved with more DM particles. \citet{sellwood+1998} showed that the less well resolved a galaxy is, the easier it is for the disc to thicken. As TNG50 has a relatively large softening length we expect the discs may be somewhat thicker than those in nature. On the other hand, how severe this effect is might depend on the gas content of the disc, since kinematic heating of the gas disc may be radiated away, modulating its effect. Detailed investigation of this however is beyond the scope of this work.

Possible effects in cosmological simulations that may \emph{encourage} buckling include strong bars possibly due to overly efficient removal of gas from the centers of the galaxies producing more radially anisotropic distributions at the center. However, we find no evidence that TNG50 produces excessively strong bars. In Fig.~\ref{fig:Mstar_versus_Rbar}, we show the ($\logMstar,\Rbar$)-plane and overlay in blue circles observational data from the Spitzer Survey of Stellar Structure in Galaxies \citep[S$^4$G,][]{sheth+2010}, taken from \citet{Erwin_2018}. This is a distance- and magnitude-limited sample.  Here, the bar radius is its deprojected semi-major axis length. We exclude galaxies with $\logMstar<10.0$ to match the TNG50 sample. From this figure, it is clear that at $z=0$, TNG50 does not produce excessively long (and by implication strong) bars compared to the observations. In TNG50 at $z=0$, median $\Rbar = 3.5_{-0.8}^{+1.5} \kpc$ after the exclusion of bars smaller than $2.6 \kpc$, and $3.0_{-0.8}^{+1.6} \kpc$ without excluding small bars (16th and 84th percentiles). The observational sample has bar lengths $4.0_{-1.0}^{+2.1} \kpc$ after the exclusion of bars smaller than $2.6 \kpc$, and $2.7_{-1.1}^{+2.3} \kpc$ for all bars.

Recent work by \citet{frankel_etal_2022} concluded that TNG50 bars appear in general to be too short. This would contribute to the suppression of buckling in TNG50, yet about half ($49$ per cent) of its BPs (at least for $\Rbar>2.6$ kpc) are formed through this mechanism. Had buckling been more prevalent in TNG50, and all 45 non-BPs in the plateau region ($\logMstar\geq10.5$ at $z=0$) undergone strong buckling to be classified as BCK, the buckling fraction would be 97/151 or 64\%. Had they formed a BP through WNB mechanisms, the buckling fraction would be 52/151 or 34\%. We thus estimate that in nature, the fraction of BPs observed in the current epoch which had a strong buckling episode in their past is $49\pm15$ per cent. A higher percentage is more likely assuming some buckling suppression in TNG50.

Moreover, the fact that we find 7 galaxies, or $4_{-1}^{+2}$ per cent of the barred sample have buckled recently (within $1.5 \Gyr$ of $z=0$) implies a significant probability of detecting recently buckled galaxies in the local Universe. Indeed, eight candidates have so far been identified \citep{erwin_debattista16,li+17,xiang+2021}.


\section{Summary}
\label{s:summary}

We have examined $\numbars$ barred galaxies with $\logMstar \geq 10.0$ and $\Rbar\geq2.6 \kpc$ from the TNG50 run of the IllustrisTNG simulation suite at $z=0$ for the presence of box/peanut (BP) bulges. We have analysed the differences between those barred galaxies which have BPs at $z=0$ and those which do not.

To find BPs, as well as inspection of density plots and unsharp images, we have used the fourth-order Gauss-Hermite moments of the vertical velocity distributions along the bar's major axis. We use a metric developed from its profile to measure a BP's `strength' (akin to how well defined it is), and from this, the time of BP formation. We also track the time of bar formation, and the time of major buckling, if that occurs. Our main results are:

\begin{enumerate}

\item We identify $\numbps$ BPs among $\numbars$ barred galaxies at $z=0$ ($55$ per cent). Of these, $\numbuck$ (49\%) have buckled strongly at some stage in their evolution, and $\numnonbuck$ (51\%) have not, presumably forming their BPs via resonant trapping, and/or weak buckling (Section~\ref{s:stats_z0}).

\item We find that bar characteristics (specifically length and strength) have the greatest discriminatory power in TNG50 in determining whether a barred galaxy has a BP or not, though this is in tension with observations (where stellar mass is the most important factor). The stronger and the longer the bars, the greater the fraction of them which host BPs (Section~\ref{s:logistic_regression_fits}).

\item We find the BP fraction amongst barred galaxies to be a strong function of stellar mass. TNG50 reproduces the overall shape and turning point ($\logMstar\sim10.1$) of the distribution when compared with observations. It does not reproduce the observed fraction above $\logMstar\sim10.5$, with TNG50 underproducing BPs at high mass (Section~\ref{s:stats_z0}).

\item Our analysis shows that those galaxies destined to have BPs at $z=0$ have a more quiescent history than non-BP barred galaxies even when controlled for mass (consistent with the Auriga simulations). Their last major mergers occurred  $\sim1.75\Gyr$ earlier than non-BP galaxies even when controlling for mass, and the latter also experienced many more minor mergers after this point. At $z=0$, on average, the BP galaxies have median $f_\mathrm{acc}\sim6.3$ per cent versus $11.3$ per cent in mass-matched controls (Section~\ref{ss:impact_mergers}).

\item Galaxies destined to have BPs at $z=0$ have thinner discs in relative terms, longer and stronger bars since $z\sim1$, and their discs are more radially anisotropic compared to non-BP galaxies, even when controlling for mass. Their bars are $\sim 25$ per cent stronger and $\sim 40$ per cent longer than bars in non-BP galaxies at $z=0$. BPs require a host galaxy with low anisotropy ($\sigzsigR$) to form (Section~\ref{ss:kinematics_evol}).

\item In agreement with earlier observational studies, we find that the gas fraction has little discriminatory power as to whether a galaxy has a BP at $z=0$. Within the bar, the gas fraction is so low that it has no influence on whether a BP forms or not (Section~\ref{s:gas}).

\item In TNG50, the low BP fraction at low mass is a consequence of bar youth, a manifestation of downsizing. Their youth means the bars are weaker, resulting in lower $\sigma_R$ and higher $\sigzsigR$, suppressing BP formation (Section~\ref{ss:BP_frac_discussion}).

\item Higher mass non-BP galaxies have more past mergers than lower mass non-BP galaxies, suppressing BP formation. Heating from mergers, particularly at high mass, may be excessive in TNG50 compared to nature, resulting in the plateaued profile of the distribution of the BP fraction below that in observations (Section~\ref{ss:BP_frac_discussion}).

\item Our analysis suggests that buckling may be suppressed in TNG50, even though we find 49\% of $z=0$ BPs having buckled in the past (excluding short bars). This implies that in nature, more than half (perhaps up to 65\%) of $z=0$ BP bulges may have had a strong buckling episode in the past (Section~\ref{ss:buckling_implications}).

\end{enumerate}


\section*{Acknowledgements}

We thank the anonymous referee for a thoughtful and constructive report, which has helped to improve the clarity of the paper. We also thank Annalisa Pillepich and Dylan Nelson for helpful discussions which improved the paper. SGK is supported by the Moses Holden Studentship. The IllustrisTNG simulations were undertaken with compute time awarded by the Gauss Centre for Supercomputing (GCS) under GCS Large-Scale Projects GCS-ILLU and GCS-DWAR on the GCS share of the supercomputer Hazel Hen at the High Performance Computing Center Stuttgart (HLRS), as well as on the machines of the Max Planck Computing and Data Facility (MPCDF) in Garching, Germany.
Analysis of TNG50 galaxies was carried out on Stardynamics, a 64 core machine which was funded from Newton Advanced Fellowship \#~NA150272 awarded by the Royal Society and the Newton Fund.
VC is supported by ANID-2022 FONDECYT postdoctoral grant no. 3220206.
The analysis in this paper made use of the \texttt{PYTHON} packages  \texttt{JUPYTER, NUMPY,PYNBODY} and \texttt{SCIPY} \citep{jupyternotebooks, numpy,pynbody, scipy}. Figures in this work were produced using the \texttt{PYTHON} package \texttt{MATPLOTLIB} \citep{matplotlib}.


\section*{Data availability}

The data from the IllustrisTNG suite of simulations including TNG50 can be found at the website: \url{https://www.tng-project.org/data} \citep{nelson_etal_2019}.




\bibliographystyle{mnras}
\bibliography{TNGBP.bbl} 



%
\appendix
%
%
%
%


\section{Sample Selection}
\label{s:sample_selection}

We describe here how barred and BP galaxies were selected from TNG50.

\subsection{Barred galaxies}
\label{ss:bar_sample}

Disc galaxies were selected at $z=0$ following the method used by \citet{zhao+2020} with TNG100: we select those galaxies with stellar masses $M_\star \ge 10^{10} \Msun$ within a spherical radius of $30\kpc$.
We align the galaxy onto the $(x,y)$-plane using the angular momentum vector as described in Section~\ref{s:TNG50_overview}. Disc galaxies are those with stellar kinematics dominated by ordered rotation using the $\Krot$ parameter \citep{sales_etal_2010}:
\begin{equation}
\Krot = \frac{1}{M} \sum_k{ \frac{m_k v_{\phi, k}^2} {v_{x,k}^2 + v_{y,k}^2 + v_{z,k}^2} },
\end{equation}
where the sum is over all stellar particles $k$ within the $30 \kpc$ sphere, and $M$ is the total stellar mass within the sphere. $\Krot$ measures the fraction of stellar kinetic energy committed to in-plane rotation.
Disc galaxies are defined as those having $\Krot \ge 0.5$ \citep{zhao+2020}. These cuts result in a sample of $\numdiscs$ disc galaxies at $z=0$. We refer to this as the `disc sample'. For this sample, the number of stellar particles within $10 \Reff$ is $1.7-96\times10^5$ (median $4.0\times10^5$) and the number of gas particles ranges from just $221$ to $2.0\times10^6$ (median $2.1\times10^5$) at $z=0$. From this point on, metrics are computed within a spherical radius of $10R_\mathrm{eff}$ unless otherwise noted.

Since a bar is a bisymmetric deviation from axisymmetry, we define the global bar strength, $\Abar$, as the amplitude of the $m=2$ Fourier moment of the stellar particle surface density distribution, projected onto the $(x,y)$-plane.
We calculate the global bar amplitude as:
\begin{equation}
A_{\mathrm{bar}} = \left|\frac{\sum_k m_k e^{2i\phi_k}}{\sum_k m_k}\right|,
\end{equation}
where $\phi_k$ and $m_k$ are the azimuthal angle and mass of each star particle. We compute the radial profile of a bar's amplitude as:
\begin{equation}
a_2(R) = \left|\frac{\sum_{k,R} m_k e^{2i\phi_k}}{\sum_{k,R} m_k}\right|,
\end{equation}
where now the sum runs over all stellar particles within a given cylindrical annulus of radius $R$. We also calculate the phase of the $m=2$ Fourier moment within each annulus. Within the region where the $m=2$ phase is constant to within $10 \degrees$, we define the maximum of $a_2(R)$ as $\Atwomax$. We consider a galaxy to be barred if $\Atwomax>0.2$. This results in a sample of $\numbarsinclshort$ barred galaxies ($44$ per cent of the disc sample).

Many methods have been devised for measuring bar lengths, each having advantages and disadvantages \citep[e.g.][]{agu_etal_00, athanassoula_misiriotis02, erwin05, michel_dansac+wozniak_2006,cuomo_et_al_2021}. At each time step in the simulation for each barred galaxy identified above, we compute the bar's radius, $\Rbar$, as the average of the cylindrical radius at which the amplitude of the $m=2$ Fourier moment reaches half its maximum value (so $\Atwomax/2$) after its peak, and the cylindrical radius at which the phase of the $m=2$ component deviates from a constant by more than $10\degrees$ beyond the peak in $a_2(R)$ (the median of half the difference between the two measures is $\sim 20$ per cent of the bar radius).

We restrict our attention to those BPs whose extent is at least twice the softening length ($288\pc$ at $z=0$) to be secure in our recognition of BPs. In their study of 84 moderately inclined local barred galaxies \citetalias{erwin_debattista17} found that the range of sizes of the BP as a fraction of the bar radius is $0.25-0.76$ \citep[see also][]{lutticke+00}. To account for BPs of similar relative sizes, we require a minimum bar radius of $\sim2 \kpc$, and so we exclude galaxies with $\Rbar < 2 \kpc$. We argue in Section~\ref{ss:refine_rbar_cut} for a more stringent cut of $\Rbar < 2.6 \kpc$. This more stringent cut eliminates 75 of the $\numbarsinclshort$ barred galaxies, resulting in a final sample of $\numbars$ barred galaxies out of $\numdiscs$ ($\sim31$ per cent) disc galaxies.

We refer to this as the `barred sample'. We emphasize that it includes only those galaxies with bars at $z=0$, whose radius is $\ge2.6 \kpc$, having $\Atwomax>0.2$, and not those which may have formed bars which dissolved (or shrank to a radius $<2.6\kpc$ or weakened to below $\Atwomax=0.2$) before $z=0$.

We note that \citet{rosas_guevara+2022} (RG22) also studied barred galaxies in TNG50 with $\logMstar \geq 10$, with the total stellar mass defined as that enclosed within $10 \Reff$ as in this study. Using the circularity parameter $\varepsilon=J_z/J(E)$ (where $J_z$ is the specific angular momentum of the particle around the $z$-axis, and $J(E)$ is the maximum specific angular momentum possible at the specific binding energy of a stellar particle), disc particles were defined as those with $\varepsilon\geq0.7$, and disc galaxies were defined as those with disc/total mass fraction $>50$ per cent. Using this method, and a different bar length cutoff criteria than used in this study (RG22 used a bar radius limit based on the softening length), they found 349 discs and 105 (30\%) bars, compared to this study's $\numdiscs$ discs and 191 (31\%) bars.

The existence of a longer and stronger bar can perturb stars efficiently to dynamically hotter orbits. This mechanism can gradually reduce the mass fraction of stars with high $\varepsilon$. It is thus not surprising that the criterion used in RG22 found a smaller number of disk galaxies and barred galaxies. In \citet{du+2020,dumin+2021}, a careful decomposition of galaxies with $\Krot>0.5$ showed that the mass ratios of their kinematically-defined disk structures are larger than 50\% of the total stellar mass. We have also visually confirmed that the galaxies we selected using $\Krot$ have clear disk structures. Therefore, we retain the use of $\Krot$ in disc galaxy selection.

\begin{figure}
  \includegraphics[width=\hsize]{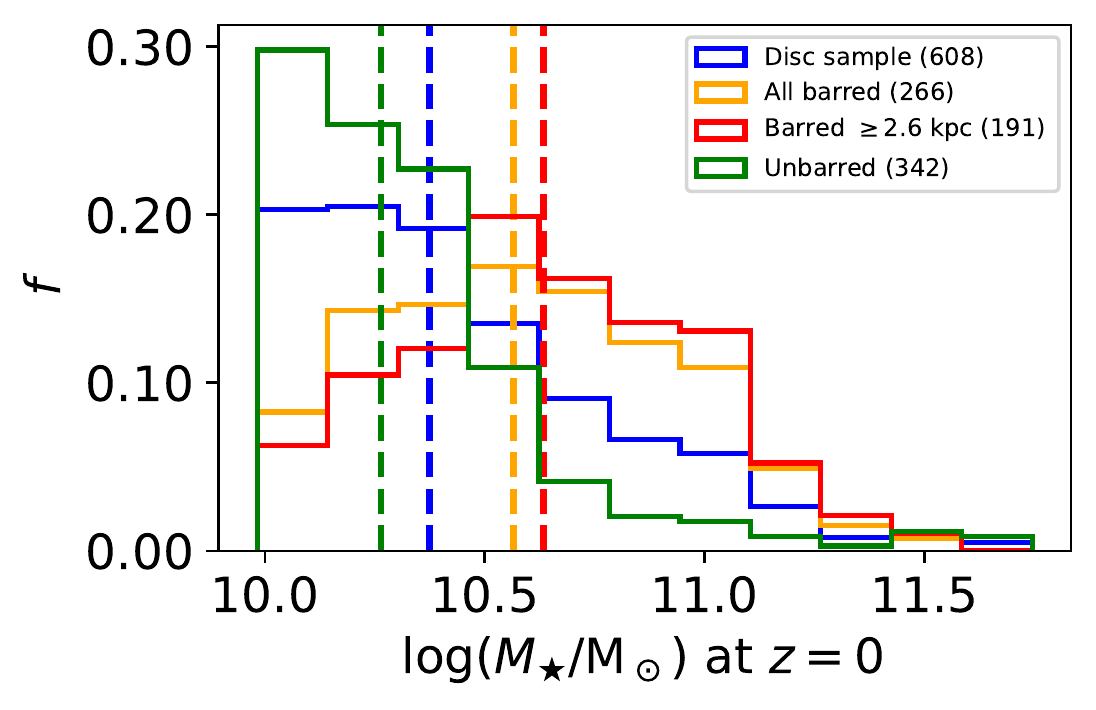}
  \caption{The distribution of stellar mass at $z=0$ for all disc galaxies (disc sample -- $\numdiscs$ galaxies, blue line), for all galaxies in the disc sample with bars ($\Atwomax>0.2$, $\numbarsinclshort$ galaxies, orange line) and for the galaxies in the disc sample without bars ($\numnonbars$ galaxies, green line) at $z=0$. The red line shows the distribution for all bars within the disc sample with $\Rbar\geq2.6 \kpc$ ($\numbars$ galaxies, see Section~\ref{ss:bar_sample}). The vertical dashed lines show the median stellar masses in each group.}
  \label{fig:M_star_dist_z0}
\end{figure}

Fig.~\ref{fig:M_star_dist_z0} shows the stellar mass distributions at $z=0$ for the disc ($\numdiscs$ galaxies), all barred ($\numbarsinclshort$ galaxies), barred with $\Rbar \ge 2.6 \kpc$ ($\numbars$ galaxies) and unbarred galaxies. The stellar mass distribution of the barred sample is similar to the mass distribution of all ($\ie$ including those with $\Rbar<2.6 \kpc$) barred galaxies.

\subsection{Time of bar formation}
\label{ss:assessment_t_bar}

\begin{figure}
  \includegraphics[width=\hsize]{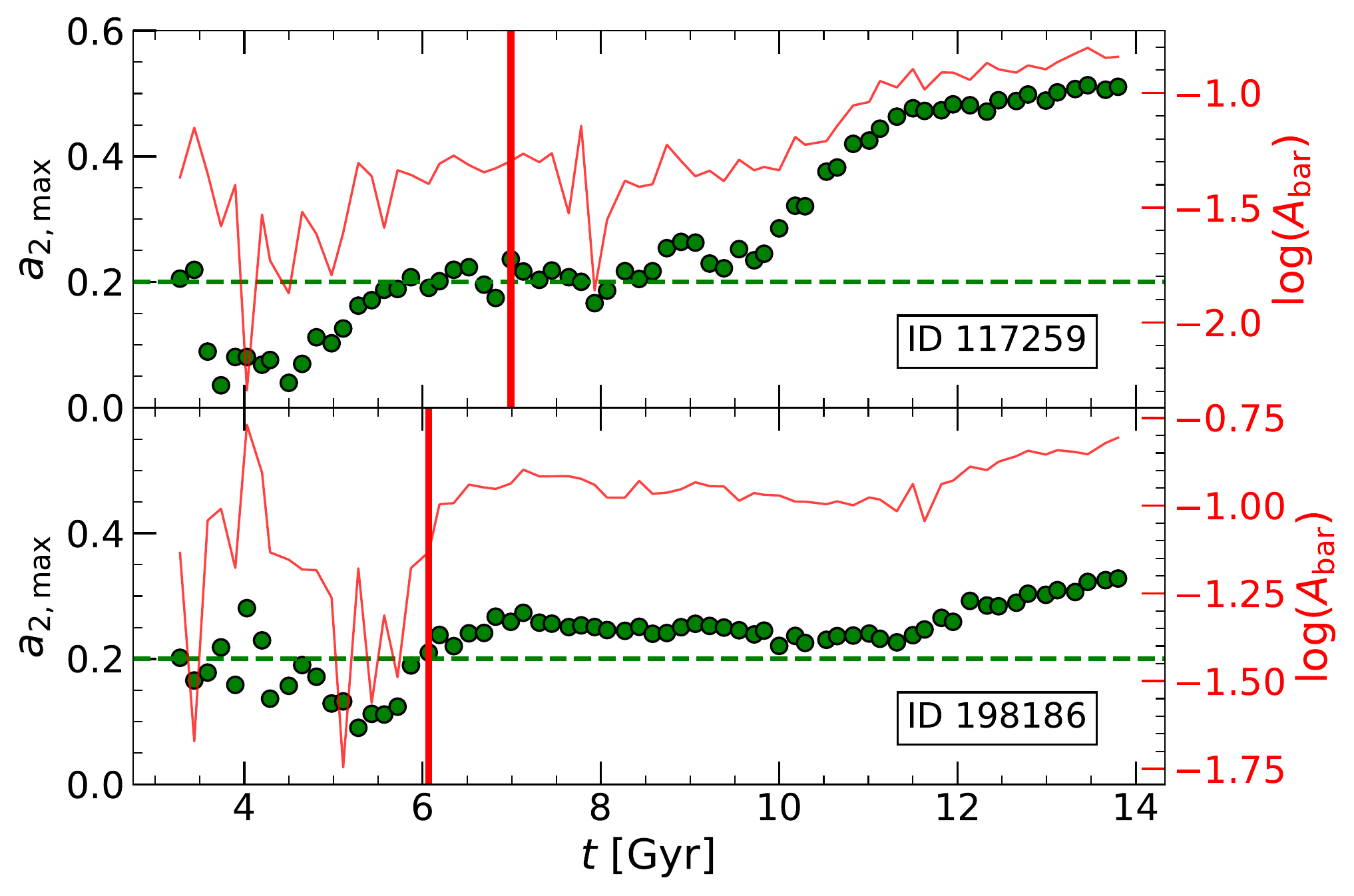}
  \caption{Bar formation time identification for two TNG50 barred galaxies. The horizontal green dashed line shows the $\Atwomax$ threshold and the green points show $\Atwomax$ values at each timestep. The solid red line shows $\log(\Abar)$ at each step. The thick red vertical lines indicate the time of bar formation, defined in Section~\ref{ss:assessment_t_bar}. The galaxies are labelled with their TNG50 Subhalo IDs.}
  \label{fig:bar_form_time}
\end{figure}

Following \citet{algorry_etal_2017}, we track the evolution of the $m=2$ Fourier amplitude for the barred sample, and set the threshold of bar formation at $\Atwomax = 0.2$. We consider the bar to have formed when we detect at least four consecutive snapshots with $\Atwomax>0.2$, taking the formation time as the first step where this occurs. We denote this time as $\tbar$. Once this is identified, we compare with the evolution of $\Abar$. Since during bar formation the bar amplitude grows exponentially, after the instability has saturated we expect to see, at most, a slower secular growth. Therefore, we inspect the evolution of $\Abar$ for each galaxy as a check on the reasonableness of the derived $\tbar$. Examples of this procedure are shown in Fig.~\ref{fig:bar_form_time}, where $\Abar$ stabilises within two or three snapshots after $\tbar$. This Figure also shows that bars can continue to grow after $\tbar$.

This method works well in 161 ($84$ per cent) of the barred galaxies, as confirmed by visual inspection of the density profiles. However, in 30 ($16$ per cent) barred galaxies, the evolution of $\Abar$ and $\Atwomax$ is too noisy to find $\tbar$ within the algorithm's rules. In these cases, we inspect the ($x, y$) density distributions at each redshift and set $\tbar$ manually. Note that 14 (7\%) of the bars have $\zbar$ earlier than $z=2$; in the middle panel of Fig.~\ref{fig:gas}, the only place where this matters, we exclude these galaxies from the analysis.

\subsection{Identification of BPs}
\label{ss:id_BPs}

We now describe our method for identifying BP bulges in barred galaxies at $z=0$. We emphasize that we do not consider galaxies which may have had BPs at earlier epochs that do not survive to $z=0$. Hence we do not study the evolution of the BP fraction as a function of redshift.

\begin{figure}
  \includegraphics[width=\hsize]{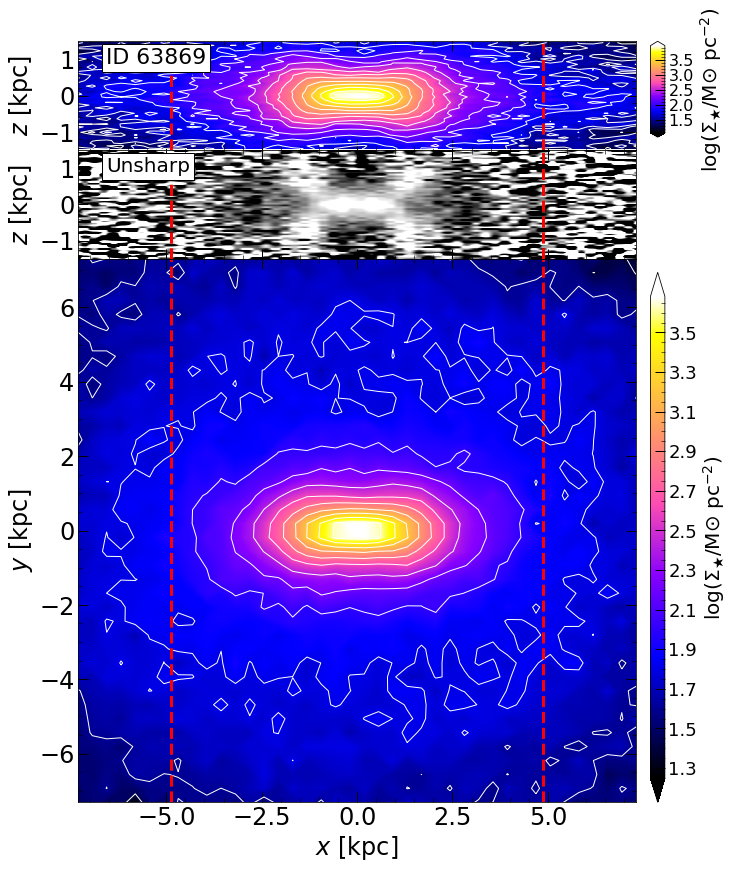}
  \caption{Example of ($x, z$)- and ($x, y$)-plane stellar density distributions (middle and bottom panels) and unsharp mask of the ($x, z$) distribution (top) for the barred galaxy {\bf 63869} at $z=0$. The bar radius is indicated by the vertical red dashed lines. In the ($x, z$) projections, a cut of $|y|<0.3\Rbar$ is used. The `X' shape of the BP bulge is apparent in the unsharp mask and pinching is visible in the ($x, z$)-plane density contours.}
  \label{fig:example_density_plots}
\end{figure}

Starting with each barred galaxy at $z=0$, we align the bar along the $x$-axis using the two-dimensional inertia tensor of the stellar density distribution in the ($x, y$)-plane. The tensor is computed within a cylindrical radius of $\Rbar$ ($\Rbar$ is found beforehand without the need to align the bar).
We then identify BPs visually initially, and use a strength metric described in Section~\ref{ss:BP_strength} for validation, and to provide a quantitative measure which we can track through each galaxy's evolution. We examine density maps in the ($x, z$) (edge-on) plane, as well as unsharp masks of the edge-on density distribution. We use a cut of $|y|<0.3\Rbar$ to maximize the visibility of any BP component present whilst maintaining a reasonable particle count. The upper two panels of Fig.~\ref{fig:example_density_plots} show an example of the ($x, z$)-plane density plot and unsharp mask for one of the galaxies in the barred sample. In the bottom panel we also present the $(x, y)$ surface density, where the bar extent is clearly visible. Note the pinched contours in the ($x, z$)-plane, which produces an `X'-shape in the unsharp mask, indicating the presence of a BP.

\subsection{Quantification of BP strength}
\label{ss:BP_strength}

The strength of a BP is a challenging concept to define, but is important for our analysis. Previous methods of determining the BP strength have included using the median height \citep[\eg][]{fragkoudi+2020}, mass excesses when modelling the bulge as a spheroidal component \citep[\eg][]{abbott+17} and $m$-fold deviations from a pure ellipse in edge-on views \citep[\eg][]{ciambur+2021}.

In this work we measure the fourth order Gauss-Hermite moment of the vertical velocity distributions \citep{gerhard93,vdm_fra_93} along the bar's major axis. The fourth order term in this series, $h_4$, measures how peaked the distribution is compared to a Gaussian. If the distribution has $h_4>0$ at a particular point, then it is more sharply peaked than a Gaussian. If $h_4<0$ then the distribution is flatter than a Gaussian. \citet{debattista+05} demonstrated that a BP bulge, when viewed face-on, produces two deep minima in the $h_4$ profile, on either side of the galactic centre along the bar major axis. This was confirmed observationally by \citet{mendez-abreu+08} in NGC~98. The minima appear because the vertical velocity distribution of stars is broadened by the presence of a BP. As did \citet{debattista+05}, in addition to the $h_4$ profile, we also examined the fourth order Gauss-Hermite moment of the height distribution along the bar major axis ($d_4$), but found this to be noisier and more challenging to constrain than $h_4$. Thus for our BP strength metric we rely solely on $h_4$.

We calculate the $h_4$ profile of $v_z$ along the bar major axis (we continue to use a cut of $|y|\leq0.3R_{\mathrm{bar}}$) for each barred galaxy at every redshift, and identify the minima in $h_4$ on either side of $x=0$, if present.
A high signal-to-noise (S/N) ratio is required to compute accurate Gauss-Hermite moments. To achieve this, we bin along the bar major axis with an adaptive number of bins (found by experimentation) based on the particle count within the cut. We compute the S/N ratio as $\sqrt{N_p}$, where $N_p$ is the particle count within each bin \citep{dumin+16}. If we find minima in the resulting $h_4$ profile, we broaden the bins until we reach S/N $\ge50$ at the location of the minima.
To aid in detecting the peaks and valleys, we smooth the resulting $h_4$ profiles using a second order Butterworth filter \citep{butterworth_1930} and interpolate using a cubic spline. We use the \texttt{SIGNAL} module of the \texttt{PYTHON} \texttt{SCIPY} library to search for peaks and valleys in the profiles, setting a limit $\lambda$ on the number of extrema per kpc in the profile, rejecting any profile which has too many peaks or valleys per kpc as being too noisy. After some experimentation we set $\lambda=1.75\kpc^{-1}$. A profile without a valley on each side of $x=0$ is deemed to be a non-BP profile. We avoid valley locations too close to the galactic centre (within $10$ per cent of the bar radius from the centre), and too close to the end of the bar (within $10$ per cent of the end). Selecting the deepest valley on each side of $x=0$, we compute their mean, and denote this as $\overline{h_{4,{\rm valley}}}$. We take the mean of the peak in $h_4$ between these two valleys (in many galaxies this frequently consists of just one broad peak), denoted as $\overline{h_{4,{\rm peak}}}$, and use the difference as the peak$-$valley amplitude $\mathcal{B}$:
\begin{equation}
    \mathcal{B} = \overline{h_{4,{\rm peak}}} - \overline{h_{4,\rm{valley}}}.
\end{equation}
We use this dimensionless quantity as the BP strength metric, akin to its prominence. Its growth indicates large vertical excursions along the bar region, \ie\ a BP bulge. We use the term `prominence' as one can observe BP bulges which are large in physical size (radius) but appear as a weakly defined peanut, or smaller BPs in radius but with strongly defined `X' shapes. So $\mathcal{B}$ measures how well-defined the BP is. Hereafter, we use the term `BP strength' throughout, on the understanding that this is a strength metric based on kinematics. \citet{debattista+05} demonstrated a strong correlation between the 4th-order moment of the velocity and density, so we are satisfied with its suitability as a measure of BP strength. Uncertainties on $\mathcal{B}$ are computed using the differences between the raw and smoothed $h_4$ profiles at the valleys and peak, added in quadrature.

Many bars without BPs have no detectable valleys in their $h_4$ profiles, and therefore have $\mathcal{B}=0$.

\subsubsection{Buckling versus weak/non-buckling BPs}
\label{ss:assessment_buckling}

It is generally accepted that BPs can form via two principal mechanisms. The first is the most morphologically obvious -- the buckling instability \citep{raha+91, merritt_sellwood94, martinez-valpuesta_shlosman04, erwin_debattista16, smirnov_sotnikova19}, a large deviation from vertical symmetry, followed by a rapid rise in vertical thickness in the inner regions of the bar. The second pathway is a BP which grows via the resonant trapping of stars, albeit more gradually \citep{combes_sanders81, combes+90, quillen02, sellwood_gerhard_2020}. This can be difficult to distinguish from weak buckling. (Recently, \citet{li_shlosman+2023} demonstrated a link between overlapping planar and vertical 2:1 resonances and the buckling instability, providing a link between resonant excitation and buckling. This hints that the two pathways to a BP may not be as distinct as previously thought, although as we show, buckling is very evident in the evolution of $\mathcal{B}$.)

It is possible that some galaxies can form and grow their BP bulge through a mixture of buckling and resonant trapping which is difficult to disentangle. We distinguish those BP bulges which experienced a strong buckling episode in their history and those which did not.  In isolated simulations, buckling happens rather rapidly, in $\sim0.5-1$ Gyr \citep[e.g.][]{martinez-valpuesta_shlosman04, martinez-valpuesta+06,lokas_2019, cuomo_etal_2022, li_shlosman+2022}. This seems consistent with observations \citep[e.g.][]{erwin_debattista16}. The minimum, maximum, and median difference between snapshots in TNG50 from $z=2$ to $0$ are 87, 236 and 159 Myr, respectively, thus providing good temporal resolution for detecting strong buckling. We note that strong buckling is obvious in $(x, z)$ density distributions and unsharp mask images. It is accompanied by a strong signature in the $h_4$ profile (two deep valleys appear on either side of $x=0$), and hence a rapid increase in the BP strength, $\mathcal{B}$.

\begin{figure*}
  \includegraphics[width=\hsize]{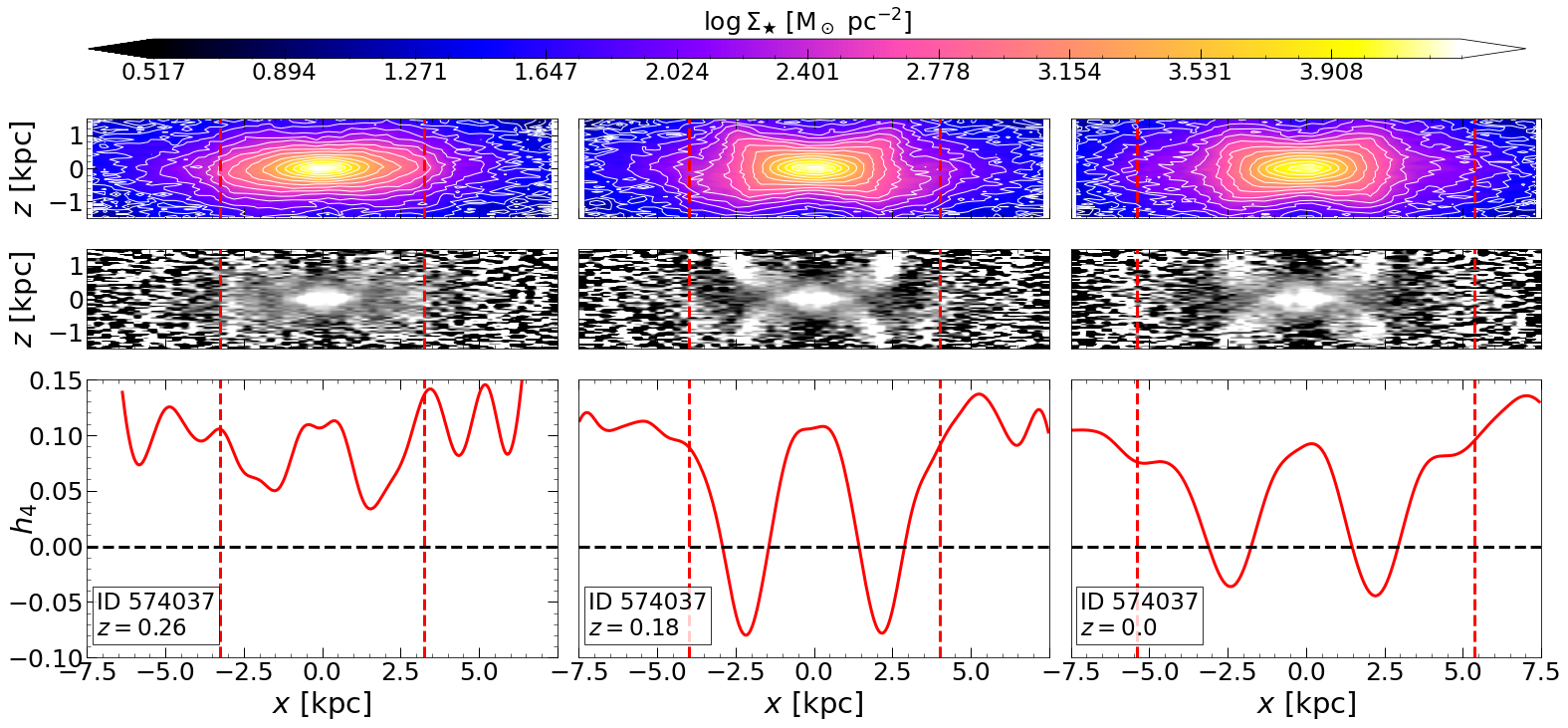}
  \caption{Example of a buckling galaxy, Subhalo ID {\bf 574037}, before, at and after its buckling redshift, $z=0.18$. Top row: stellar surface density in the $(x, z)$-plane. Middle row: unsharp mask of the surface density in the $(x, z)$-plane. Bottom row: smoothed $h_4$ profiles along the bar major axis. Columns represent, from left to right, $z=0.26, 0.18, 0$. The bar radius is indicated by the vertical red dashed lines. The black dashed line shows $h_4=0$ for reference. Each panel uses the cut $|y|<0.3\Rbar$.}
  \label{fig:buckling_galaxy}
\end{figure*}

For each galaxy with a BP at $z=0$, we examine unsharp mask images and density plots, noting if vertical asymmetry in the $(x, z)$-plane surface density map appears suddenly at any time, alongside the development of an asymmetric `X' shape in the unsharp masks. If it does, and this is accompanied by the prompt formation of two deep valleys in the $h_4$ profile, and a steep rise in the evolution of $\mathcal{B}$, then we deem the galaxy to have experienced strong buckling. We label the time at which this occurs as $\tbuck$ and denote these galaxies as the `BCK sample'. An example is shown in Fig.~\ref{fig:buckling_galaxy} where the vertical asymmetry of galaxy {\bf 574037}, at the time of buckling, stands out in the stellar density, and the `X' shape is visible in the unsharp mask. Note the evolution of the $h_4$ profile (bottom row) where we see the clear formation of two deep valleys on either side of $x=0$ at buckling. After buckling, there is a reduction in the depth of the valley in $h_4$.

We investigated the use of other buckling indicators, such as $A_\mathrm{buck}$ \citep{debattista+06} and $A_{1z}$ \citep{li_shlosman+2023}, but found these to be very noisy in TNG50, unlike in isolated simulations. In TNG50, we found a sharp increase in $\mathcal{B}$ (see the example in the top left panel of Fig.~\ref{fig:evol_h4_examples}) to be a much more reliable indicator of buckling, because it is insensitive to small misalignments of galaxy inclination (up to $\sim 30\degree$).

While strong buckling is obvious, we may miss examples of weak buckling, possibly classifying BPs which had weak (possibly recurrent) buckling episodes in their history as having formed via resonant capture. We define those BPs where we did not detect strong buckling as the weak/non-buckling (`WNB') sample.

\begin{figure*}
  \includegraphics[width=\hsize]{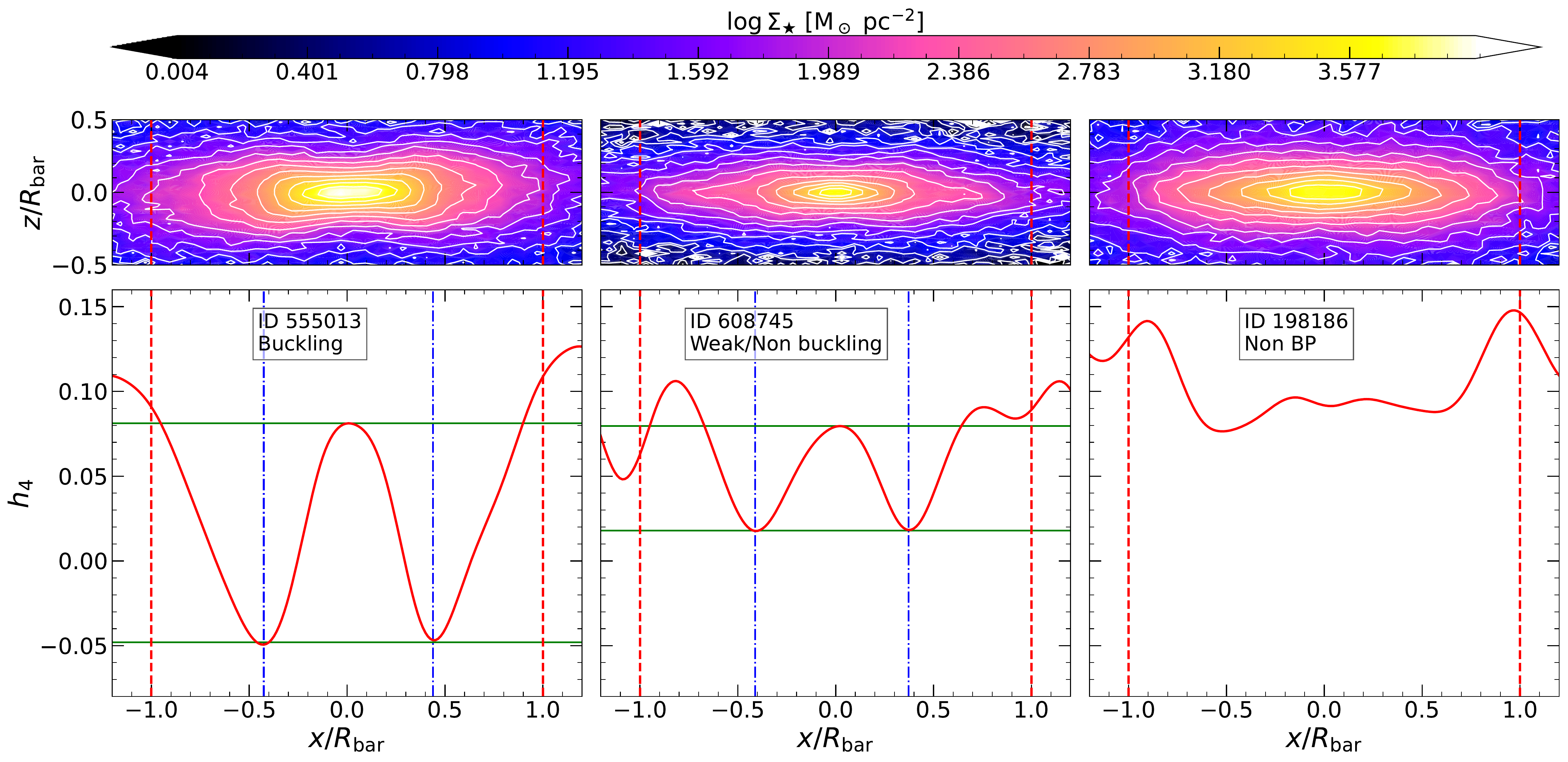}
  \caption{$h_4$ profiles along the bar major axis (bottom panels) for three galaxies at $z=0$: a buckled BP galaxy (left column), a galaxy with a BP but which has no major buckling episode (middle column) and a barred galaxy without a BP (right column). The top panels show the stellar surface density in the $(x, z)$ plane (vertical scale is shown as $z/R_\mathrm{bar}$). All panels are shown for $|y|<0.3\Rbar$. The smoothed profiles are shown in solid red lines. The bar radius is indicated by the vertical red dashed lines. The two deep minima detected by the BP algorithm (blue vertical dot-dash lines) are present in the buckling galaxy, shallower ones in the weak/non-buckling galaxy, and the galaxy without a BP has a profile with shallow valleys. The green horizontal lines represent the mean valley and peak $h_4$ levels. The galaxies are labelled with their TNG50 Subhalo IDs.}
  \label{fig:three_h4_z0_profiles}
\end{figure*}

Fig.~\ref{fig:three_h4_z0_profiles} shows three $h_4$ profiles at $z=0$ which we identify as BCK, WNB and non-BP barred galaxies. Note the deep minima with $h_4<0$ in the BP galaxy which has undergone buckling \citep[consistent with the findings of][]{sellwood_gerhard_2020}. The minima in the WNB galaxy are not as deep, but still prominent. The non-BP $h_4$ profile still has some peaks and valleys, but is considerably flatter in the central regions than the BP galaxies.

\begin{figure}
  \includegraphics[width=\hsize]{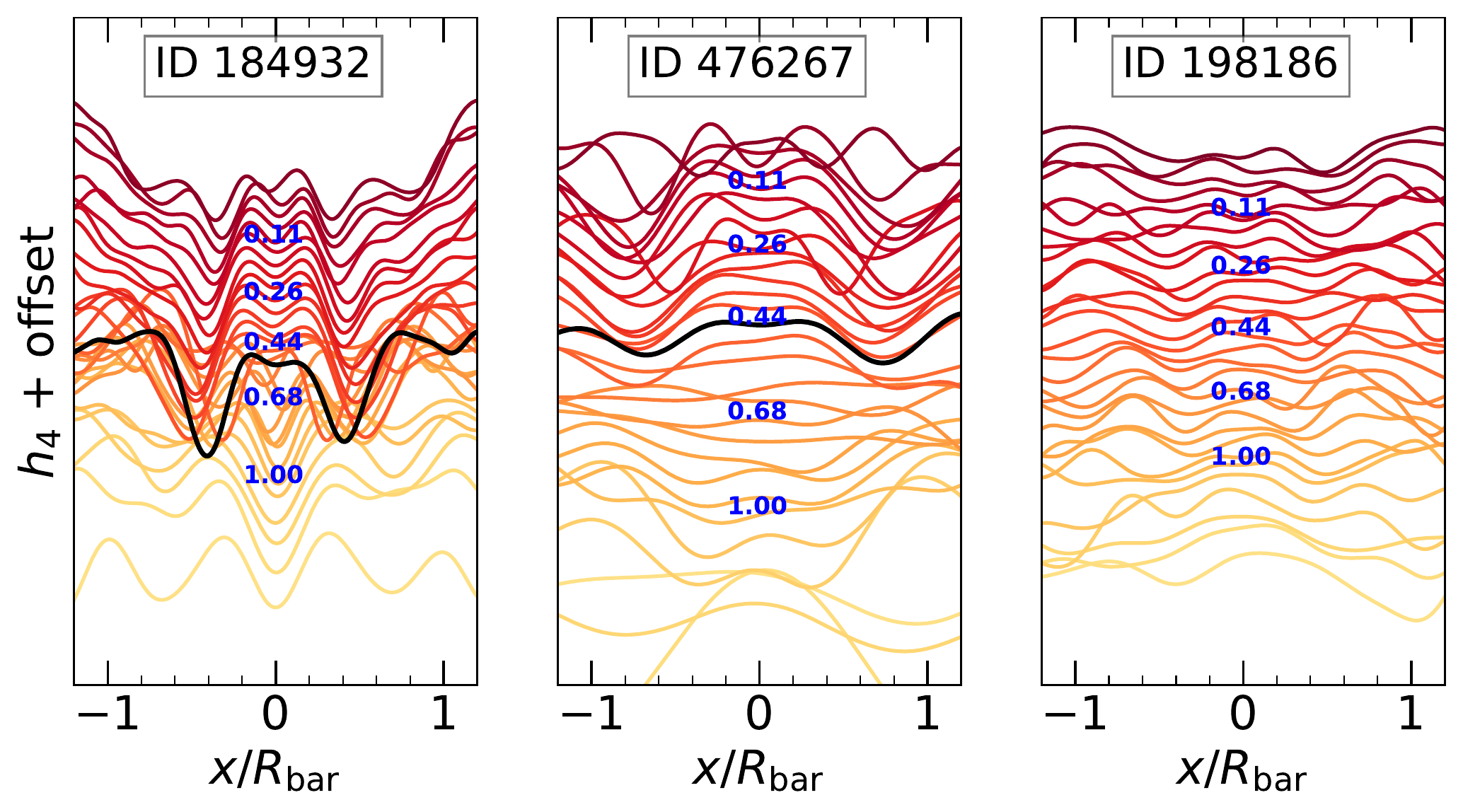}
  \caption{Examples of the evolution of the $h_4$ profiles along the bar major axis for three representative galaxies, a BCK (left panel), WNB (middle panel) and non-BP barred (right panel) galaxy. Time progresses upwards towards the darker colours, and we plot every other redshift from $z=1.5$ for clarity. We apply a constant offset to separate the profiles vertically and every fifth time step, we label the redshift in blue. The profiles in solid black lines are those at the time of BP formation (Section~\ref{ss:time_BP_formation}). The galaxies are labelled with their TNG50 Subhalo IDs.}
  \label{fig:h4_time_lapse}
\end{figure}

We illustrate the evolution of the $h_4$ profiles, from redshift $z=1.5$ to $z=0$, in Fig.~\ref{fig:h4_time_lapse} for representative BCK, WNB and non-BP barred galaxies. In the BCK and WNB galaxies, clear minima in $h_4$ are present after the BP forms, while the non-BP galaxy shows no such feature. The minima in buckling galaxies appear suddenly, and are deeper than in the WNB profiles.

\subsubsection{Refinement of the \Rbar\ cut and final barred sample}
\label{ss:refine_rbar_cut}

\begin{figure}
  \includegraphics[width=\hsize]{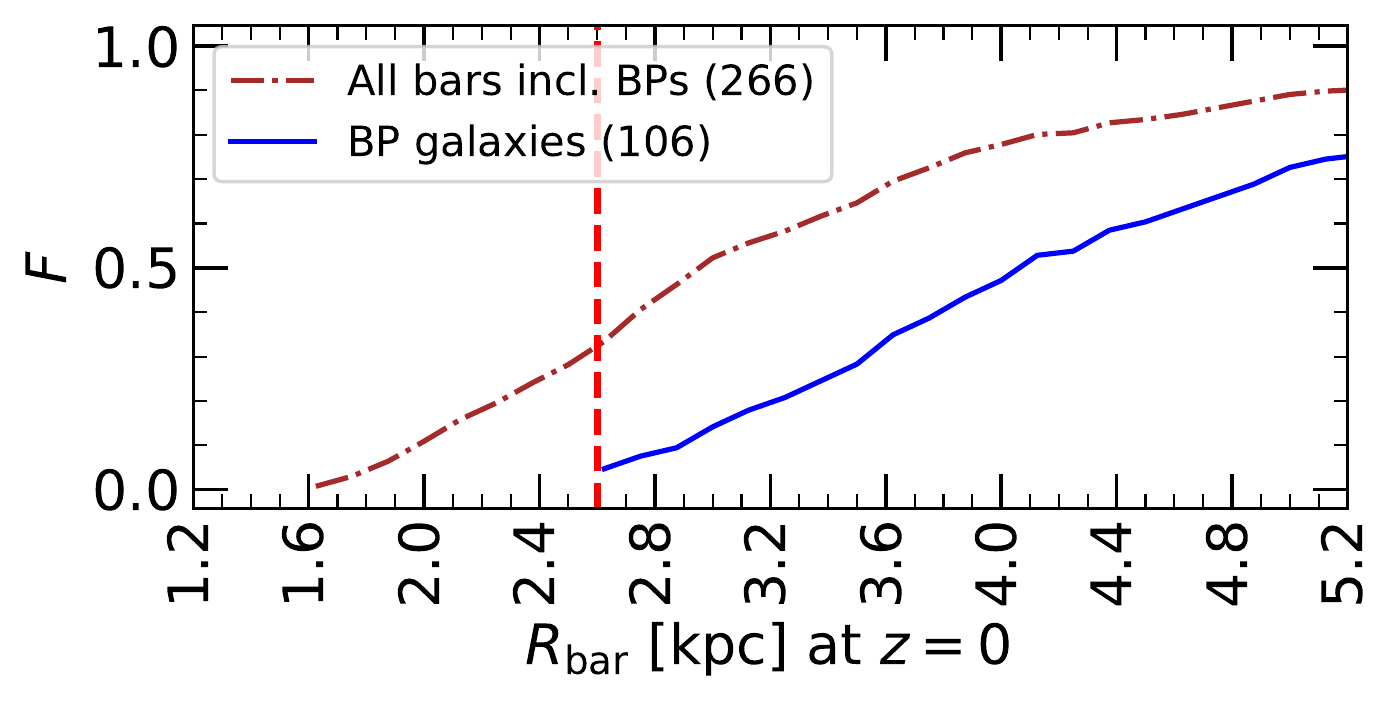}
  \caption{The cumulative distribution of $\Rbar$ for} all barred galaxies (those with $\Atwomax>0.2$ but with no cut in $\Rbar$) (brown dot-dashed line), and BP galaxies amongst them (blue solid line). To show details at small radius, the $x$-axis limit is set to $\sim5$ kpc. The vertical dashed red line marks our bar cut, $\Rbar=2.6 \kpc$.
  \label{fig:Rbar_cut}
\end{figure}

Our initial bar radius cut was $2 \kpc$, based on arguments presented in Section~\ref{ss:bar_sample}. Figure~\ref{fig:Rbar_cut} shows the cumulative distribution of $\Rbar$ for BP and non-BP galaxies, which includes all bars with $\Atwomax>0.2$. We see that a cut of $2.6 \kpc$ marks the radius where both cumulative fractions begin to rise linearly in a reasonably consistent way. Therefore, we select those bars with $\Atwomax>0.2$ and $\Rbar\ge2.6\kpc$ as the barred sample. Compared to using a cut at $\Rbar\ge 2\kpc$, this eliminates a further 58 barred and 5 BP galaxies.

After this cut, the visual inspection and BP strength determination results in a population of $\numbps$ BPs out of $\numbars$ galaxies in the barred sample ($\sim55$ per cent) at $z=0$. We refer to those galaxies with a BP at $z=0$ as the `BP sample', while the remaining barred galaxies without BPs at $z=0$ constitute the `non-BP sample' ($\numnonBPbars$ galaxies).

\subsection{Control sample}
\label{ss:control_sample}

To help investigate the conditions which lead some barred galaxies to form BPs and others not, we produce a sample of non-BP barred galaxies which match the stellar mass distribution of our BP sample. For each BP galaxy, we select a barred but non-BP galaxy which is closest in terms of total stellar mass at $z=0$. Although the matching is rather simple, it helps illuminate key differences in galaxy parameters, giving us insight into why those barred galaxies without BPs do not form them.

To construct the control sample, for every galaxy in the BP sample, we iterate through every galaxy in the non-BP barred sample to form combinations of each BP and a non-BP galaxy. For each combination, we calculate $\Delta M_\star/M_\star$, the fractional difference between the stellar masses at $z=0$.
These combinations are then sorted in ascending order by $\Delta M_\star/M_\star$ for matching. Since we have 84 non-BP galaxies (we exclude one which is in the process of buckling at $z=0$), and $\numbps$ BP galaxies, we must pair some controls with more than one BP. To maximize the match in stellar mass, we allow each control to be used up to twice, $\ie$ pairing is done \emph{with replacement}. 60 of 84 (71\%) unique control galaxies are paired amongst the BP galaxies. We refer to the population of control galaxies as `the control sample'.

\begin{figure}
  \includegraphics[width=\hsize]{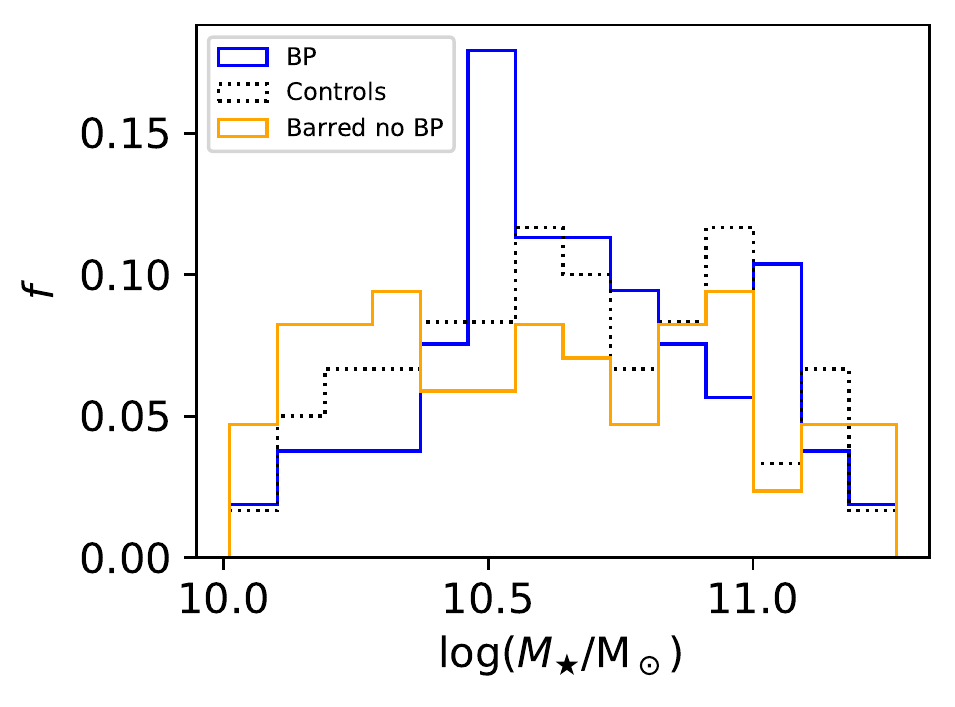}
  \caption{Distribution of $\logMstar$ for the BP (blue) and control (dotted black) samples at $z=0$ and for comparison, all barred galaxies without BPs (orange). A good match between BP and control samples is achieved with a two-sample KS test having $p=0.38$, signifying negligible difference in their distributions.}
  \label{fig:z0BPcontrols}
\end{figure}

Fig.~\ref{fig:z0BPcontrols} shows the distribution of $\Mstar$ for BP and control galaxies. The BP galaxy and control stellar mass distributions have a two-sample KS test $p$-value of $0.38$, signifying negligible differences between the two distributions. There are much larger differences between the BP sample and the population of non-BP galaxies ($p=0.008$). Using this approach, 14 controls are paired once, 46 twice. The median fractional difference in $\Mstar$ between BP galaxies and controls is $0.99$ per cent. We find $96$ per cent of BP galaxies have a fractional difference in stellar mass with their control of less than $5$ per cent. Thus, when comparing BP and control galaxies, $\Mstar$ is reasonably well controlled for.

\subsection{BP sample validation}
\label{ss:bp_samp_valid}

\begin{figure}
  \includegraphics[width=\hsize]{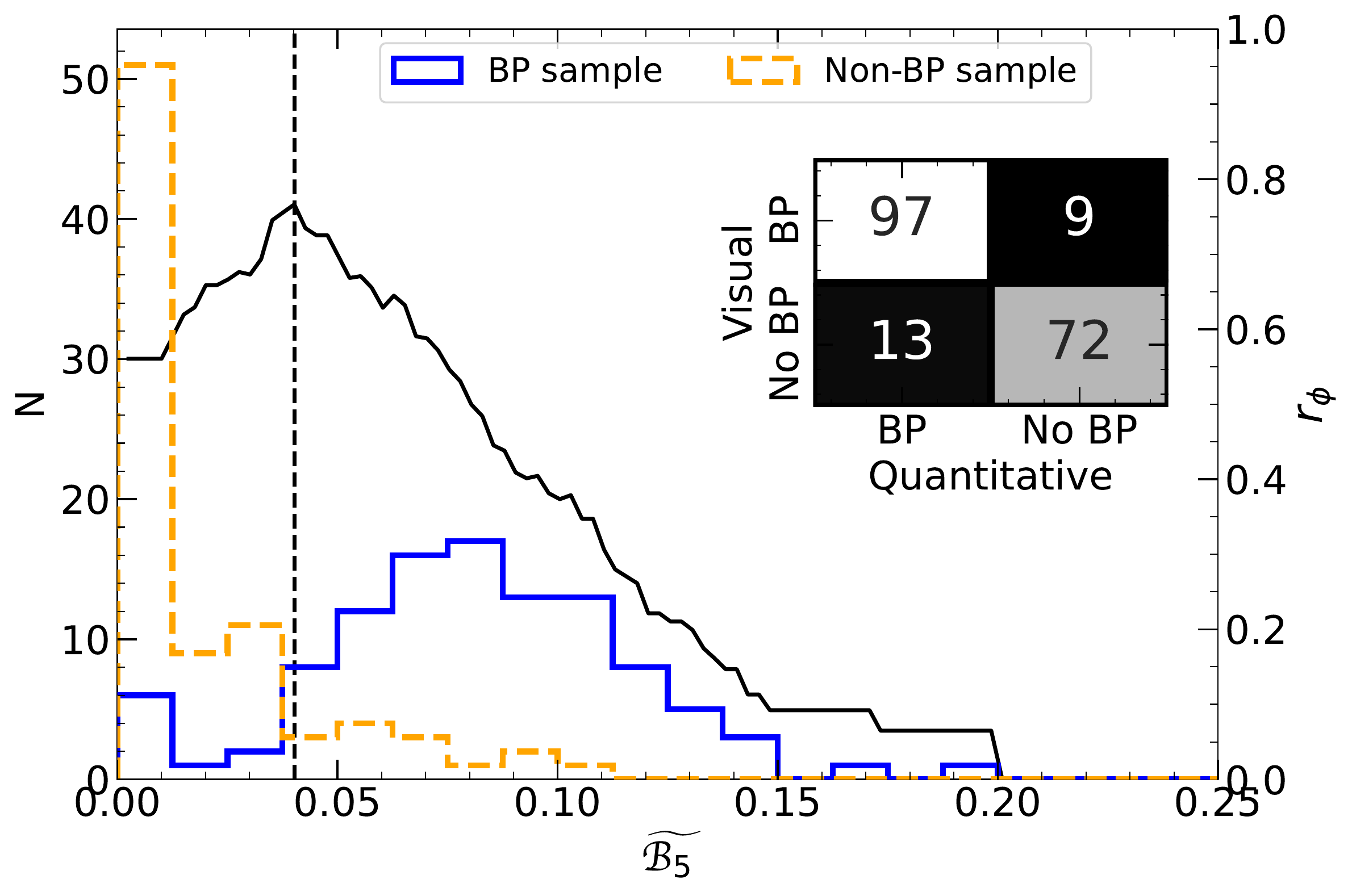}
  \caption{The distributions of the median of $\mathcal{B}$ for the last five time steps, $\widetilde{\mathcal{B}_5}$, for the BP (blue) and non-BP samples (orange) (left axis). The black line (right axis) shows $r_\phi$, the `phi coefficient' of \citet{yule_1912, matthews_1975}, a correlation coefficient between the binary classifications methods, as a function of $\widetilde{\mathcal{B}_5}$ which is used for the threshold value of BP classification. The confusion matrix in the top right corner is a comparison of the visual and quantitative classifications using the $\mathcal{B}_{\rm crit}$ (maximum $r_\phi$) value indicated by the vertical dashed line.}
  \label{fig:BPstr_final5}
\end{figure}

We use the BP strength metric, $\mathcal{B}$, to perform a quantitative test of the BP sample obtained through visual inspection, since $\mathcal{B}$ may be noisy, for example in cases of galaxies which experience interactions. We calculate the median of $\mathcal{B}$ in the final five snapshots for each galaxy, $\widetilde{\mathcal{B}_5}$, for the BP and non-BP samples and present their distributions in Fig.~\ref{fig:BPstr_final5} (left axis). As expected, non-BP galaxies have near zero values of $\widetilde{\mathcal{B}_5}$, with a tail to larger values (some galaxies have transient large values of $\mathcal{B}$). The BP sample clearly has overall larger values of $\widetilde{\mathcal{B}_5}$ than the controls, with a large range. However, the distributions overlap. Using a single critical value of $\mathcal{B}$ to classify barred galaxies as BP or non-BP would therefore introduce confusion in the classifications. We compare the visual and quantitative classifications based on applying a threshold $\mathcal{B}_{\rm crit}$, taking the visual classification to be the `truth'. We use the $\mathcal{B}_{\rm crit}$ value to make a `predictive' classification of each galaxy solely using each galaxy's value of $\widetilde{\mathcal{B}_5}$ (a BP being present if $\widetilde{\mathcal{B}_5} \ge \mathcal{B}_{\rm crit}$).

We compute the mean square contingency coefficient, $r_\phi$\footnote{$r_\phi=[(\mathrm{TP}\times\mathrm{TN})-(\mathrm{FP}\times\mathrm{FN})]/\sqrt{(\mathrm{TP}+\mathrm{FP})(\mathrm{TP}+\mathrm{FN})(\mathrm{TN}+\mathrm{FP})(\mathrm{TN}+\mathrm{FN})}$, where TP~$=$~number of true positives, TN~$=$~number of true negatives, FP~$=$~number of false positives and FN~$=$~number of false negatives for each value of $\mathcal{B}_{\rm crit}$ assessed.}, a correlation coefficient between binary classification methods \citep{yule_1912, matthews_1975}, as a function of $\mathcal{B}_{\rm crit}$, then finding the value which results in the closest match to the visual classification. The $r_\phi$ value for each threshold value is also presented in Fig.~\ref{fig:BPstr_final5} (right axis, black line). We find that a critical value of $\widetilde{\mathcal{B}_5} = \Bcrit$ results in an $r_\phi$ value of $0.735$, and an overall accuracy (number of true positives and negatives as a fraction of all classifications) of $86.9$ per cent at reproducing the visual classification. The confusion matrix using this critical value is presented in the top right of Fig.~\ref{fig:BPstr_final5}. This demonstrates that the $\mathcal{B}$ metric captures the BP component of the bulge well, with some uncertainty in the individual values as well as uncertainty in our visual classifications. While comparing BP classifications between authors gives minimal disagreement, this test demonstrates the challenge of identifying BPs with an automated routine even in simulations.

\subsection{Measuring the time of BP formation}
\label{ss:time_BP_formation}

We use the BP strength $\mathcal{B}$ to determine the time of BP formation. We calculate $\mathcal{B}$ at each time step, and set a threshold of $\mathcal{B}_{\rm crit}=\Bcrit$ found in Section~\ref{ss:bp_samp_valid}, considering that a BP is present when $\mathcal{B}>\mathcal{B}_{\rm crit}$. We find the earliest time step with no more than four subsequent consecutive time steps with $\mathcal{B}$ below the threshold (or no value for $\mathcal{B}$). We consider this to be the time when the BP formed, which we denote as $\tbp$. The constraint on the reasonably contiguous nature of $\mathcal{B}$ avoids early transient signals (perhaps caused by interactions) and spurious signals caused by poor alignment. It sets $\tbp$ at an epoch around which a BP is firmly established. That is, a few `missing' time steps with $\mathcal{B}<\Bcrit$ would not cause us to discount a time step being $\tbp$, unless this condition were protracted.

\begin{figure}
  \includegraphics[width=\hsize]{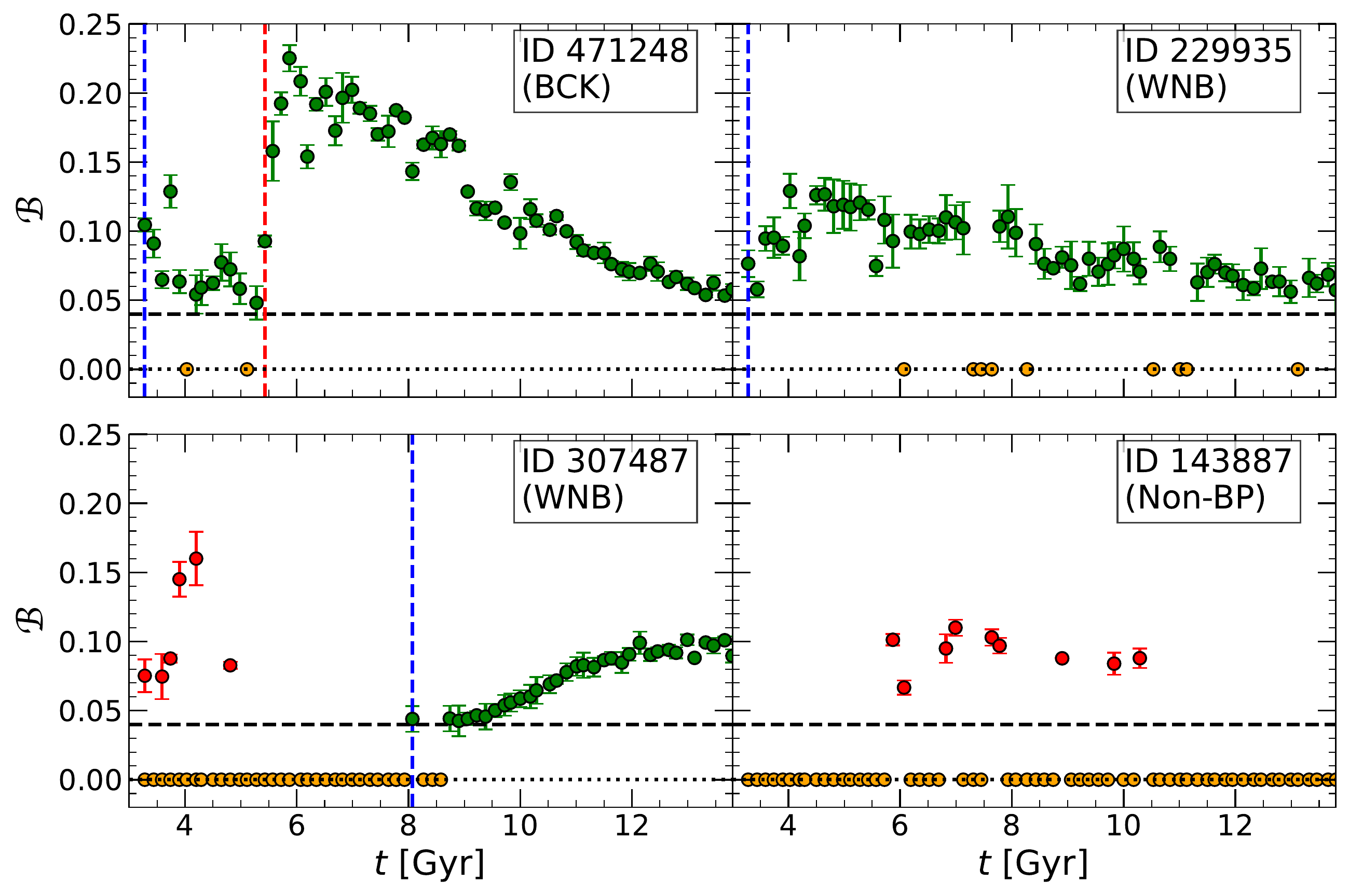}
  \caption{BP strength $\mathcal{B}$ for four galaxies, one buckling, two WNB and one non-BP galaxy. The black dashed horizontal line shows $\mathcal{B}_{\mathrm{crit}}$. Green points are those which meet our contiguity and threshold requirements for a BP, red those which do not (Section~\ref{ss:time_BP_formation}). The blue vertical dashed line marks $\tbp$, the time of BP formation (by definition, the non-BP galaxy has no BP at $z=0$). The red dashed vertical line marks the time of buckling. The redshifts where the algorithm calculates no $\mathcal{B}$ value are shown by the orange points along the $x$ axis, and we mark $\mathcal{B}=0$ with a dotted black horizontal line. The galaxies are labelled with their TNG50 Subhalo IDs.}
  \label{fig:evol_h4_examples}
\end{figure}

Fig.~\ref{fig:evol_h4_examples} shows examples of the evolution of $\mathcal{B}$ for an example buckling (BCK) galaxy, two WNB galaxies and a non-BP galaxy, with their identified $\tbp$. We mark in green those points sufficiently contiguous in $\mathcal{B}$ (and larger than $\mathcal{B}_{\mathrm{crit}}$) for us to consider them a suitable measure of BP strength. 
We see a sudden, rapid increase in $\mathcal{B}$ at $\tbuck$ followed by a period of approximately linear decline in the buckled galaxy whose BP formed $\sim1\Gyr$ before buckling. The decline in $\mathcal{B}$ is typical of buckling galaxies in TNG50, but varies across the sample. \citet{sellwood_gerhard_2020} noted an initially negative $h_4$ value immediately after buckling in their $N$-body models, which increased to around zero by the end of the evolution. The evolution of $\mathcal{B}$ in the BCK galaxies is consistent with their result.

The evolution of $\mathcal{B}$ in the WNB sample is more varied, as exemplified by the two WNB panels. Some exhibit an evolutionary pattern in $\mathcal{B}$ similar to buckling galaxies (although in the unsharp masks and density we were unable to see evidence of any strong buckling in them), others show a steady increase in $\mathcal{B}$ with time. Still others show a steady decrease in $\mathcal{B}$ after an initial rise at $\tbp$, highlighting the difficulty in disentangling weak buckling from resonant capture-built BPs. We also verify the reasonableness of $\tbp$ by visually inspecting the $(x,z)$ density distributions and unsharp masks, ensuring a BP is discernible at this time.

For the non-BP galaxies, there is usually no pattern in the evolution of $\mathcal{B}$, and often no value could be calculated owing to a lack of a substantial $h_4$ peak$-$valley structure. Three controls do have some periods of contiguous evolution of $\mathcal{B}$ before $z=0$, but $\mathcal{B}<\mathcal{B}_\mathrm{crit}$ at $z=0$. They have indications of having had BPs in the past, but no longer at $z=0$. Five controls have contiguous values of $\mathcal{B}>\mathcal{B}_{\mathrm{crit}}$ up to $z=0$ but no discernible BPs in the density plots and unsharp masks.

We check the consistency of $\tbp$ and $\tbar$, checking those galaxies where $\tbp$ was more than 1 Gyr before $\tbar$. This occurs in 3 galaxies and we reinspect these galaxies' $h_4$ and density plots. We amend $\tbp$ forward manually a few time steps in these cases.

%

\bsp	
\label{lastpage}
\end{document}